\documentclass[referee, iicol, sn-mathphys]{sn-jnl}
\usepackage{amsmath}
\usepackage{amsfonts}
\usepackage{hyperref}
\usepackage{graphicx}
\graphicspath{Figures/}
\usepackage{algorithm}
\usepackage{algpseudocode}
\usepackage[caption=false]{subfig}
\definecolor{ao(english)}{rgb}{0.0, 0.5, 0.0}
\definecolor{awesome}{rgb}{1.0, 0.13, 0.32}

\usepackage{xcolor}

\begin{document}

\title[Article title]{Wake-induced response of vibro-impacting systems}


\author[1]{\fnm{Rohit} \sur{Chawla}} \email{rohit.chawla@ucd.ie} 

\author[1]{\fnm{Aasifa} \sur{Rounak}} \email{aasifa.rounak@ucd.ie} 

\author[2]{\fnm{Chandan} \sur{Bose}} \email{chandan.bose@ed.ac.uk} 

\author*[1]{\fnm{Vikram} \sur{Pakrashi} \email{vikram.pakrashi@ucd.ie}} 


\affil[1]{\orgdiv{UCD Centre for Mechanics, Dynamical Systems and Risk Laboratory, School of Mechanical and Materials Engineering}, \orgname{University College Dublin}, \orgaddress{\city{Dublin}, \country{Ireland}}}



\affil[2]{\orgdiv{School of Engineering, Institute for Energy Systems}, \orgname{University of Edinburgh}, \orgaddress{\city{Edinburgh}, \country{Scotland}}}


\abstract{The stability and bifurcation behaviour of a wake-induced vibro-impacting oscillator is studied. The effects of a discontinuity on the stability of the structure while it is undergoing phase-locked motions due to the surrounding fluid-structure interactions (FSI) are examined. The primary structure and the near wake dynamics are modelled as a harmonic oscillator and a Van der Pol oscillator, respectively, and are weakly coupled to each other via acceleration coupling.  Qualitative changes in the dynamical behaviour of this system are investigated in the context of discontinuity-induced bifurcations (DIBs) that result from the interaction of fluid flow and non-smoothness from the primary structure. Phenomenological behaviours like the co-existence of attractors and period-adding cascades of limit cycles separated by chaotic orbits are observed. The existence of these phenomena is demonstrated via stability analysis using Floquet theory and the associated Lyapunov spectra. In addition, the behaviour of orbits in the local neighbourhood of the barrier is defined using a higher-order transverse discontinuity map. This mapping is implemented to obtain the respective Lyapunov exponents. Solutions obtained using this modified algorithm are demonstrated to accurately predict both stable and chaotic regimes, as observed from the corresponding bifurcation diagrams.}.


\maketitle

\section{Introduction}

\begin{table}[!]\label{tab A}
\caption{Description of variable names and symbols.}
    \begin{tabular}{@{}llll@{}}
    \toprule
    Variable name & Symbols & Units\\
    \midrule
    \textbf{Structure variables} & &\\
    Displacement of structure & $Y$ & m\\
    Diameter of structure & $D$ & m\\
    Mass of oscillator & $m$ & kg\\
    Mass of structure & $m_s$ & kg\\
    Fluid-added mass & $m_f$ & kg\\
    Stiffness of oscillator $Y$ & h & N $m^{-1}$\\
    Damping of oscillator & $\varrho$ & Ns $m^{-1}$\\
    Damping of structure & $r_s$ & Ns $m^{-1}$\\
    Fluid-added damping & $r_f$ & Ns $m^{-1}$\\
    Time & $\tau$ & s\\
    Lift force & $S$ & N\\
    Natural frequency of structure & $\Omega_s$ & $s^{-1}$\\
    Fluid-added mass coefficient & $C_m$ & -\\
    Fluid flow velocity & $U$ & m$s^{-1}$\\
    Density of fluid medium & $\rho$ & kg $m^{-3}$\\
    \midrule
    \textbf{Wake variables} & &\\
    Reduced vortex lift coefficient & $X$ & -\\
    Vortex shedding frequency & $\Omega_f$ & $s^{-1}$\\
    Nonlinear parameter & $\epsilon$ & -\\
    Coupling force of structure on wake & $F$ & N\\
    Vortex lift coefficient & $C_L$ & -\\
    Reference lift coefficient & $C_{Lo}$ & -\\
    Strouhal number & $S_t$ & -\\
    \midrule
    \textbf{Dimensionless variables} & &\\
    Structure displacement & $y$ & -\\
    Barrier location & $\sigma$ & -\\
    Reduced vortex lift coefficient & $x$ & -\\
    Stroboscopic displacement & $y_0$ & -\\
    Time & $t$ & -\\
    Ratios of oscillator frequency & $\delta$ & -\\
    Damping ratio & $\xi$ & -\\
    Stall parameter & $\gamma$ & -\\
    Mass ratio & $\mu$ & -\\
    Reduced coupling force on wake & $f$ & -\\
    Coupling parameter & $A$ & -\\
    Reduced fluid-flow velocity & $U_r$ & -\\
    Reduced lift force & $s$ & -\\
    Reduced coupling variable & $M$ & -\\
    Coefficient of restitution & $r$ & -\\
    \midrule
    \textbf{Piecewise-smooth variables} & & \textbf{Dims.}\\
    Generalized state & $\mathbf{x}$ & $\mathbb{R}^{n}$\\
    Generalized vector field & $\mathbf{F}(\mathbf{x})$ & $\mathbb{R}^{n}$\\
    Perturbation to $\mathbf{x}$ & $\mathbf{y}$ & $\mathbb{R}^{n}$\\
    Discontinuity constraint & $H(\mathbf{x})$ & $\mathbb{R}^{n - 1}$\\
    Discontinuity barrier & $\mathbf{\Sigma_2}$ & $\mathbb{R}^{n - 1}$\\
    Poincar\'e section & $\mathbf{\Sigma_1}$ & $\mathbb{R}^{n - 1}$\\
    State $\mathbf{x}$ at impact & $\mathbf{x}_i$ & $\mathbb{R}^{n}$\\
    Reset map & $\mathbf{R}(\mathbf{x}_i)$ & $\mathbb{R}^{n}$\\
    Perturbed state at impact & $\mathbf{x}_0$ & $\mathbb{R}^{n}$\\
    Perturbed state on $\Sigma_2$ & $\mathbf{x}_2$ & $\mathbb{R}^{n}$\\
    Reset perturbed state on $\Sigma_2$ & $\mathbf{x}_3$ & $\mathbb{R}^{n}$\\
    Mapped perturbed state & $\mathbf{x}_4$ & $\mathbb{R}^{n}$\\
    Instant of impact state $\mathbf{x}_i$ & $t_i$ & $\mathbb{R}^{1}$\\
    Flight time to impact state $\mathbf{x}_2$ & $\delta_1$ & $\mathbb{R}^{1}$\\
    Perturbation during impact & $\mathbf{y}_-$ & $\mathbb{R}^{n}$\\
    Mapped perturbation & $\mathbf{y}_+$ & $\mathbb{R}^{n}$\\
    Saltation matrix & $\mathbf{S}_i$ & $\mathbb{R}^{n \times n}$\\
    \botrule
    \end{tabular}
\end{table}

Understanding the impact dynamics between two structures subjected to vortex-induced vibrations is a crucial aspect from the standpoint of structural safety, structural integrity, and longevity of various engineering systems, especially in environments where multiple structures are closely spaced and their interactions under fluid flow can lead to complex dynamic behaviours. Structures in proximity can interfere with each other’s flow patterns, leading to complex coupled vibrations that are not present when structures are isolated \cite{xue2015practical,raiola2016wake}. Understanding these interactions is vital for predicting and mitigating potentially destructive resonant conditions. In a recent numerical study, De \textit{et al.} \cite{de2022vortex} investigated the unsteady flow dynamics of a circular cylinder colliding with the wall. A penalty-type collision model based on the shortest distance between the cylinder and the wall was employed in this study. Two distinct natures of impact and gracing motion of the cylinder over the response branches were reported. Hu \textit{et al.} \cite{hu2024model} recently proposed a two-dimensional vessel-riser array mathematical model based on the Blevins wake model to study the collision dynamics of the riser array considering various riser arrangement modes. Nevertheless, to the best of the authors' knowledge, no existing study has focused on the discontinuity-induced bifurcation scenario in detail from the purview of dynamical systems theories. The engineering systems of the present interest are offshore platforms and structures \cite{Ellinas}, marine risers operated in group \cite{tian2020static}, bridge piers and supports, and arrays of marine turbines or floating structures, among others. For example, collision prevention control for marine riser arrays is imperative in order to increase production efficiency and reduce maintenance costs. A combination of a variety of environmental loads that arise from waves, currents, and those arising from the impact of vessels shall result in the development of extreme stresses \cite{chandrasekaran2020design}. Different passive and active control methods have been proposed in the past to mitigate the impact of ocean currents on risers \cite{huse2000impulse}. The findings of this paper can inform the efficient design to avoid and control such collisions. Practical scenarios of vortex-induced vibrations subjected to vibro-impacts can be found in other offshore engineering applications as well. Some examples include cross-flow induced chaotic responses in heat exchanger tubes impacting loose supports \cite{paidoussis1992cross}, floating buoyant systems attached to risers and mooring lines undergoing impacts with a rigid harbour \cite{virgin2009some}, fatigue failure of submerged structures like large-scale trash racks of hydropower stations subjected to vibro-impacts due to clearance issues \cite{xue2023nonlinear}, floating vessels subjected to impacts due to liquid sloshing impact loads \cite{ibrahim2014recent}, vibro-impact dynamics in rolling of ships due to impacts with a rigid harbour \cite{ibrahim2014recent} during ship grounding, amongst others. Vortex-induced vibration is commonly observed in nature as well as in various bluff-body-shaped engineering structures. Although the VIV characteristics of cylindrical structures are extensively studied in the literature \cite{williamson2004vortex,williamson2008brief}, the dynamical behaviour of such systems when impacting with rigid barriers is inadequately explored. Such a scenario can arise during lifetime operations of large submerged structures like trash racks of hydro-power stations and ship mooring lines when the structure undergoes large amplitude oscillations, resulting in an impact with a nearby secondary structure. Such impacts are often due to loosened joints or lack of clearance. Despite a need to understand these dynamics, there is a paucity of literature on this subject.
Diverse approaches exist to study the statics and dynamics of engineering applications subjected to vibro-impacts. While engineers and naval designers might focus on design codes and structural assessment, dynamicists often focus on analytical approaches to determine the nonlinear dynamics of such systems for a wider range of boundary and initial conditions. This allows for establishing better insights into the nonlinear phenomena of such systems, informing the engineers of stability, safety and performance.  Large and complex structures can often be modelled by canonical equations  (e.g. harmonic oscillator, Duffing equation, Van der Pol, Mathieu, etc.) representative of the phenomena they are associated with. This paper is motivated by a similar approach and investigates the vibro-impact dynamics of structures subjected to vortex-induced vibrations. Harmonic oscillators subjected to impacts (known as impact oscillators \cite{bernardo2008piecewise}) with a rigid or deformable barrier are known to cause aperiodic and chaotic oscillations \cite{nordmark1991non}. Such systems exhibit dynamically rich bifurcation behaviours, which are not commonly observed in systems that evolve smoothly \cite{nordmark1991non}. This paper investigates various discontinuity-induced bifurcations (DIBs) due to vibro-impacts of a fluid-structure interaction system subjected to vortex-induced vibration for a wide range of fluid-flow velocities. The existing numerical methods for stability analysis, {\it i.e.} determining Floquet multipliers and Lyapunov exponents for smooth dynamical systems, need to be modified to accommodate the sudden switch in dynamical behaviour, which has been presented in detail in this study. This paper also presents a numerical methodology to determine Floquet multipliers and Lyapunov exponents, accommodating for the discontinuous transitions in the phase space, which can predict when discontinuity-induced bifurcations occur as the system is subjected to changes. The algorithm proposed in this article to estimate the Lyapunov exponents implements a higher-order transverse-discontinuity mapping \cite{chawla2022stability} that extends the more conventional methods reliant on linearized approaches \cite{bernardo2008piecewise,muller1995calculation} and making them more accurate in terms of the identification of an impact, as well as their quantification and prediction of trajectory in future.

Low-order models are widely used to study vortex-induced dynamics in structures immersed in fluids \cite{facchinetti2004coupling}. Such models have been observed to effectively capture the qualitative aspects of various phenomena exhibited by two-way coupled fluid-solid systems. On the other hand, exhaustive sets of experiments pose a challenge in isolating the effects of a single parameter while direct numerical simulation (DNS) of the Navier-Stokes equations are computationally expensive  \cite{newman1997direct}. For low-order models, similar to the approach taken in this paper, the near wake dynamics is typically described by a single variable that captures the fluctuating nature of vortex shedding and the associated periodic aerodynamic forces. This variable can be represented by the parameter of the non-dimensionalized Van der Pol oscillator that exhibits self-sustained limit-cycle oscillations (LCOs). With this low-order modelling approach in mind, the idealized approximation of a bluff body immersed in the fluid and executing VIV can be conceptualized as a single-degree-of-freedom (SDOF) elastically linear harmonic oscillator coupled with a Van der Pol oscillator\cite{facchinetti2004coupling}. 
A detailed qualitative and quantitative review of phenomenological models and experimental studies on VIV can be found in \cite{sarpkaya2004critical,gabbai2005overview,williamson2004vortex,williamson2008brief}. This link between a low-order canonical system with existing experimental validations is also a motivating factor behind the choice of the system.

Many such phenomenological models have been developed since the concept of a wake oscillator was first introduced by Bishop and Hassan \cite{bishop1964lift}. Balasubramanian and Skop \cite{balasubramanian1996nonlinear} modelled the near wake dynamics interacting with a cylindrical bluff body using a coupled Van der Pol oscillator. Later, Skop \cite{skop1997new} modified the wake oscillator by introducing a stall parameter. Other wake oscillator models, like the Rayleigh oscillator, were considered by Hartlen and Currie \cite{hartlen1970lift}, where the damping term was proportional to the velocity quadratically. Krenk and Nielsen \cite{krenk1999energy} adapted this wake oscillator by considering a quadratic velocity damping in the coupled Van der Pol system \textit{i.e.,} a combination of Rayleigh-Van der Pol wake oscillator. Recent developments of the wake oscillator model can be found in \cite{ogink2010wake,qu2020single}. Irrespective of the modeling approach it is important to note the phenomena that they intend to capture.

VIV is characterized by the phenomenon of lock-in or synchronization, which occurs when the vortex shedding frequency approaches the structure's natural frequency, resulting in large amplitude periodic oscillations. 
To accurately capture the lock-in phenomenon, modelling the coupling between the wake oscillator and the structural oscillator is crucial. Hartlen and Currie \cite{hartlen1970lift} initially considered a rather arbitrary velocity coupling between the two systems. Later, this velocity coupling was also adapted by Mureithi \textit{et al.} \cite{mureithi2000bifurcation} and Plaschko \cite{plaschko2000global}. In 2004, Facchinetti \textit{et al.} \cite{facchinetti2004coupling} proposed an acceleration coupling between the wake and structure oscillator to model the effect of the structure's motion on the wake dynamics.

Experiments conducted by Govardhan and Williamson \cite{govardhan2000modes,govardhan2004critical} reveal that the amplitude response observed on varying the fluid-flow velocity of an elastically mounted cylinder undergoing VIV comprises of three different bifurcation branches depending on the mass ratio $m^*$ (cylinder mass/fluid added mass) \cite{govardhan2000modes}. 
When the mass ratio of the FSI system is less than the critical value of $0.54$, the lower branch ceases to exist, and only a large amplitude response of the structure corresponding to the upper branch is observed for all higher values of $U_r$. Thus, the upper branch continues indefinitely, and the synchronization range extends to infinity. A comparison between the results obtained from the acceleration coupling model by Facchinetti \textit{et al.} \cite{facchinetti2004coupling} with that of the experiments by Govardhan {\it et. al} \cite{govardhan2000modes} and Khalak {\it et. al} \cite{khalak1999motions} show close qualitative agreements in the amplitude responses in this regime. Therefore, such phenomenological models can be used to qualitatively and quantitatively capture the essential physics underpinning the dynamics of VIV systems.

It is worth noting that the response dynamics of a VIV system subjected to the impacts of rigid barriers are yet to be explored. This study addresses this gap and  numerically investigates the effect of a rigid barrier obstructing the motion of a structure exhibiting VIV. A rigid barrier is considered with the assumption that the structural deformation due to such impacts is negligible, which is often the case in a range of engineering applications. 
A rigid barrier leads to abrupt interactions of motions. In the field of dynaimcal systems, such abrupt interactions are often studies considering piecewise-smooth (PWS) systems. To explore how non-smoothness impacts synchronization, regimes in which alack of barrier results in a lock-in or synchronization have been examined in this paper. 

PWS systems often exhibit contrasting bifurcation behaviour compared to smooth systems. A non-smooth transition divides the phase space into sub-spaces by what is known as the discontinuity boundary. An impact occurs when orbits in phase space interact with this boundary. A typical feature observed due to an impact is discontinuity-induced bifurcation (DIB), resulting in occurrences like grazing \cite{banerjee2009invisible}, sticking-sliding \cite{di2002bifurcations} and chattering \cite{budd1994chattering} near the discontinuity boundary \cite{bernardo2008piecewise}. DIBs often lead to period adding cascades of steady states, interspersed with aperiodic or chaotic solutions or unconventional routes to chaos \textit{i.e.}, via grazing \cite{nusse1994border,chin1994grazing,jiang2017grazing,oestreich1996bifurcation,awrejcewicz2003bifurcation,chawla2022stability}. Understanding the dynamical behaviour of coupled fluid-solid systems in the existence of a barrier is thus crucial to the design and lifetime operations of offshore structures, individually or when interacting with each other.

In this paper, a non-smooth FSI system undergoing VIV with a mass ratio below the critical value $m^* = 0.52$  is considered.
A low-order model proposed by Facchinetti \textit{et al.} \cite{facchinetti2004coupling} is selected in this paper where the structure is represented by an SDOF oscillator, while the wake is modelled using a Van der Pol oscillator, This captures the transition from the pre-lock-in lower amplitude response to the higher amplitude upper branch in the sustained lock-in regime for higher values of $U_r$.  The structure is coupled linearly to the wake variable and the effect of the motion of the structure on wake dynamics is modelled via an acceleration coupling. Non-smoothness manifests via to the presence of a rigid barrier which, upon interaction with the structure, leads to instantaneous velocity reversal modelled using a restitutive law. 
A detailed investigation of the amplitude responses of the structure in this paper allows for the visualisation of rich dynamical behaviour, such as a period-adding cascade of solutions separated by aperiodic chaotic solutions due to the presence of a non-smooth barrier. Stability analysis using Floquet theory is carried out to establish the presence of chaotic attractors \cite{nayfeh2008applied}, along with estimates of Lyapunov exponents. It is crucial to note that the numerical evaluation of the Floquet and the Lyapunov spectrum is not straightforward as compared to what can be done for smooth systems \cite{stefanski2000estimation,balcerzak2020determining,leine2012non}. The challenge lies in that fact that infinitesimally close trajectories do not interact with the border at the same time. There exists a time difference between such events, and an incorrect mapping of the state at the instant of the impact of the barrier will lead to an incorrect prediction of the state. Thus, variational equations that form the underlying basis of the above-mentioned analyses require careful consideration. The infinitesimally perturbed trajectories get mapped, at such instances, via a transverse discontinuity mapping (TDM) \cite{muller1995calculation,bernardo2008piecewise,coleman1997motions,chawla2022stability,yin2018higher}. TDM implements a corrected state transition matrix at the instant of impact, which subsequently ensures that the nearby trajectories are mapped accurately. 

This paper implements a higher-order $\mathcal{O}(2)$ TDM for transversal interactions of orbits in the local neighbourhood of the discontinuity boundary to evaluate the corresponding Lyapunov exponents. Using this approach, the Floquet and Lyapunov spectra are numerically obtained, and the corresponding bifurcation behaviour is analyzed. 

The remainder of this paper is structured as follows. Section \ref{sec 2} discusses the FSI model undergoing motion in the presence of a rigid barrier obstructing its motion. In Sec. \ref{sec 3}, the bifurcation behaviour is described. Section \ref{sec 4} discusses stability analysis. The analytical form of the TDM at impact is derived and algorithmic details have been presented here. Section \ref{sec 5} summarizes the findings of this study.

\section{Problem definition and governing equations} \label{sec 2}

\begin{figure*}
    \centering
    \includegraphics[scale = 0.8]{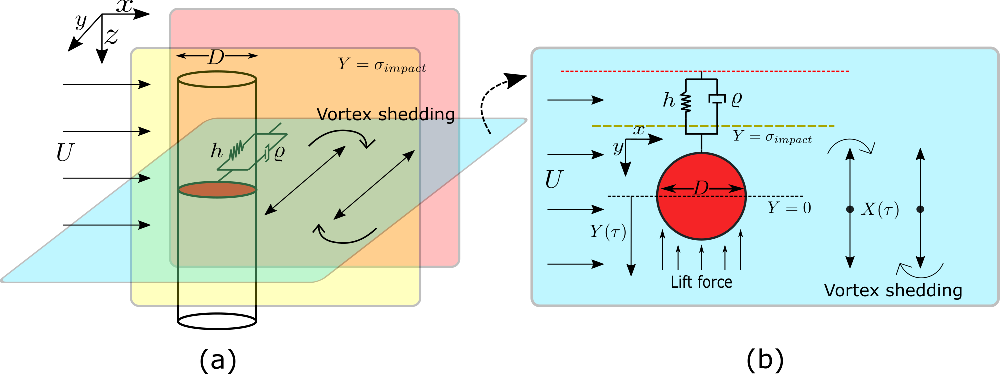}
    \caption{(a) A three-dimensional schematic representation of a cylinder of diameter $D$ with uni-directional motion in the transverse cross-flow direction $y$. A rigid barrier shown in yellow is placed at $Y = \sigma$, obstructing the motion of the cylinder executing VIV. (b) A two-dimensional schematic representation depicting the setup. The lift force exerted by the fluid on the cylinder acts in the transverse direction to the flow.}
    \label{fig 1}
\end{figure*}

Figure. \ref{fig 1} depicts a SDOF cylinder of diameter $D$ subjected to a cross flow velocity $U$. The structure exhibits limit-cycle oscillations due to its interaction with the surrounding wake. The motion of the structure is modelled as a linear oscillator $Y(\tau)$, and the periodic lift coefficient is modelled using a Van der Pol oscillator $X(\tau)$ as given by Eq. \eqref{eq 1}.
\begin{align} \label{eq 1}
    m Y'' &+ h Y + \varrho Y' = S, \\
    X'' &+ \Omega_f^2 X + \epsilon \Omega_f (X^2-1)X' = F \nonumber.
\end{align}
\noindent The prime denotes a derivative with respect to the dimensional time $\tau$. The mass of the oscillator $m$ is composed of the mass of the cylinder $m_s$ and the fluid-added mass $m_f$ \cite{blevins1977flow}, defined as $m = m_s + m_f$. The fluid-added mass is a function of the added mass coefficient $C_m$, density of fluid $\rho$ and $D$, defined as $m_f = C_m \rho D^2 \pi/4$. The damping coefficient of the oscillator $\varrho$ comprises damping effects due to the structure $r_s$ and fluid-added damping $r_f$. The fluid-added damping is a function of stall parameter $\gamma$, $\rho$, $D$ and frequency $\Omega$, defined as $r_f = \gamma \Omega \rho D^2$. In this case, where the structure is undergoing VIV due to cross-flow, $\Omega$ is the frequency of the periodic lift force $\Omega_f$. The natural frequency of the structure is defined as $\Omega_s^2 = h/m$ where $h$ is the stiffness of the oscillator. $S$ depicts the lift force acting on the structure due to its interaction with the cross-flow fluid. The nonlinearity of the Van der Pol oscillator is governed by $\epsilon$, and $F$ denotes the effects of the structure's motion on the wake dynamics.

A two-way coupling low-order model, proposed by Facchinetti \cite{facchinetti2004coupling}, is implemented in Eq. \eqref{eq 1} that represents the interactions between the cylindrical structure subjected to vortex-induced vibrations with the fluctuating wake. During vortex-induced vibrations, a lift force is exerted on the cylindrical structure at near-wake dynamics. This lift force is periodic in nature, and hence, a Van der Pol oscillator is chosen as a possible candidate for modelling purposes. The external force $S$ acting on the cylinder is only due to vorticity in the wake. The vortex-lift coefficient $C_L$ only represents the contribution of oscillating wake on the structure. Therefore, the vortex-lift coefficient $C_L$ (proportional to the lift force by definition) is represented by the Vander Pol oscillator $X = 2C_L/C_{Lo}$ where $C_{Lo}$ is the reference lift-coefficient acting on a fixed cylindrical structure during vortex-shedding. Next, the parameters for the Van der Pol oscillator \textit{i.e.}, $A/\epsilon$, are chosen that best match with experimental results. By defining a vortex-lift magnification $K = X/2$, Facchinetti analytically compared the solution of $K$ versus structure amplitude $y_0$ obtained experimentally \cite{bishop1964lift,king1977vortex,griffin1980vortex,pantazopoulos1994vortex,vickery1964flow} and showed that a value of $A/\epsilon = 40$ best fits the data. Hence, this tuned Van der Pol oscillator is linearly coupled to the structure's motion (represented by a harmonic oscillator).\\
Next, the effect of the structure's motion on the near-wake dynamics is chosen. Numerous coupling models have been previously investigated to match experimental observations. Hartley and Currie \cite{hartlen1970lift} considered a velocity coupling \textit{i.e.}, $F \propto Y'$ while Krenk and Nielsen \cite{krenk1999energy} considered a displacement coupling \textit{i.e.}, $F \propto Y$ based on direct energy transfer from the wake induced flow to the structure. Facchinetti \cite{facchinetti2004coupling} proposed an acceleration coupling \textit{i.e.}, $F \propto Y''$ that corresponds to experimental results \cite{bishop1964lift,bearman1984vortex,carberry2001forces} where the structure's vibrational response is in phase with the lift force for lower values of fluid-flow velocity $U$ and out of phase for higher values of $U$. The acceleration coupling \cite{facchinetti2004coupling} also corresponds well with experimental investigations conducted by Govardhan and Williamson \cite{govardhan2000modes} on the extent of lock-in for varying cylinder mass. Only the acceleration coupling could successfully model the presence of sustained oscillations for higher values of $U$ when the mass of the cylinder was lower than a critical mass ratio obtained by Govardhan and Williamson \cite{govardhan2000modes} experimentally. A physical insight of the acceleration coupling \textit{i.e.}, $F \propto Y''$ on the wake oscillator can be interpreted as a spring-mass oscillator with its frame of reference applied with a base excitation. Essentially, the Van der Pol oscillator is subjected to inertial coupling as if it were connected to a moving structure.\\
Using the non-dimensional mass ratio $\mu = m/\rho D^2$, the Strouhal number $S_t = \Omega_f D/2 \pi U$, the damping ratio $\xi = r_s/2 m \Omega_s$, dimensionless displacement of the structure $y = Y/D$, dimensionless time $t = \Omega_f \tau$, and  $\delta = \Omega_s/\Omega_f$, $s = S/m \Omega_f D$, $f = F/D \Omega_f$, Eq. \eqref{eq 1} can be non-dimensionalized as
\begin{align}\label{eq 2}
   \ddot{y} &+ \delta^2 y + (2 \xi \delta + \frac{\gamma}{\mu}) \dot{y} = s, \\
   \ddot{x} &+ x + \epsilon(x^2-1)\dot{x} = f\nonumber.
\end{align}
\noindent The overdot denotes the derivative with respect to $t$. 
The wake variable is defined as $x = 2 C_L/C_{Lo}$, where $C_L$ and $C_{Lo}$ are the present and reference lift coefficients. $s$ is redefined as $s = M x$ where $M = C_{Lo}/16 \pi^2 S_t^2 \mu$ is a scaling factor. A reduced fluid-flow velocity is defined as $U_r = 2\pi U/\Omega_s D$ resulting in $\delta = 1/S_t U_r$. The reduced fluid-flow velocity is chosen as a bifurcation parameter, and its effects on the vibrational responses are investigated due to an impact. Effects of structural motion on the wake dynamics are modelled using an acceleration coupling \cite{facchinetti2004coupling} defined as $f = A \ddot{y}$.

When the structure interacts with a barrier located at $y(t_-) = \sigma$, it undergoes an instantaneous reversal of velocity at the instant of impact, where $t_-$ and $t_+$ denote the instants immediately before and after impact, respectively. The instantaneous velocity reversal is defined using a coefficient of restitution $r$; see Eqs. \eqref{eq 3} and \eqref{eq 4}.
\begin{align} \label{eq 3}
    y(t_-) &= \sigma, \\
    \dot{y}(t_+) &= -r \dot{y}(t_-).\nonumber
\end{align}
and,
\begin{align} \label{eq 4}
    x(t_+) &= x(t_-), \\
    \dot{x}(t_+) &= \dot{x}(t_-) \nonumber.
\end{align}
The dynamics are governed by a smooth flow defined using Eqs. \eqref{eq 2} and a discrete map at the instant of impact defined by Eqs. \eqref{eq 3} and \eqref{eq 4}. The values implemented in Eqs. \eqref{eq 2} - \eqref{eq 4} are provided in Table \ref{tab B}. The numerical values chosen correspond well with experiments \cite{facchinetti2004coupling, govardhan2004critical}. PWS systems exhibiting such behaviour fall under the class of hybrid systems. The section below discusses the effects of an impact on the motion of the structure.

\begin{table}[h!]\label{tab B}
\caption{Numerical values }
    \begin{tabular}{@{}llll@{}}
    \toprule
    Variable name & Symbols & Values\\
    \midrule
    Strouhal number & $S_t$ & 0.2\\
    Damping & $\xi$ & 0.0052\\
    Stall parameter & $\gamma$ & 0.8\\
    Reduced mass & $\mu$ & 1.19381\\
    Nonlinearity parameter & $\epsilon$ & 0.3\\
    Coupling parameter & $A$ & 12\\
    Coefficient of restitution & $r$ & 0.8\\
    Barrier location & $\sigma$ & -0.150298\\
    \botrule
    \end{tabular}
\end{table}

\section{Bifurcation analysis} \label{sec 3}
A detailed numerical investigation of Eqs. \eqref{eq 2} along with explicit calculations of the non-smooth map that arises from the instantaneous reversal described in Eqs. \eqref{eq 3} and \eqref{eq 4} is presented here. Numerical integration has been carried out using adaptive time stepping-based solver {\tt NDSolve} in {\tt Mathematica}. 
Bifurcation analysis is carried out by varying $U_r$ and $\sigma$ as the control parameters. In order to eliminate transience, the first $500$ impacts have been discarded. 
A Poincar\'e section $\Sigma$ is defined at $\dot{y} = 0$, and the subsequent 200 impacts are observed. If the trajectory interacts with the Poincar\'e section $N$ times and returns to the same plane within a tolerance of $10^{-8}$ after elapsed time T in which the orbit repeats itself, it is labelled as the period-$N$ or PN orbit.

\begin{figure*}
    \centering
    \includegraphics[scale = 0.7]{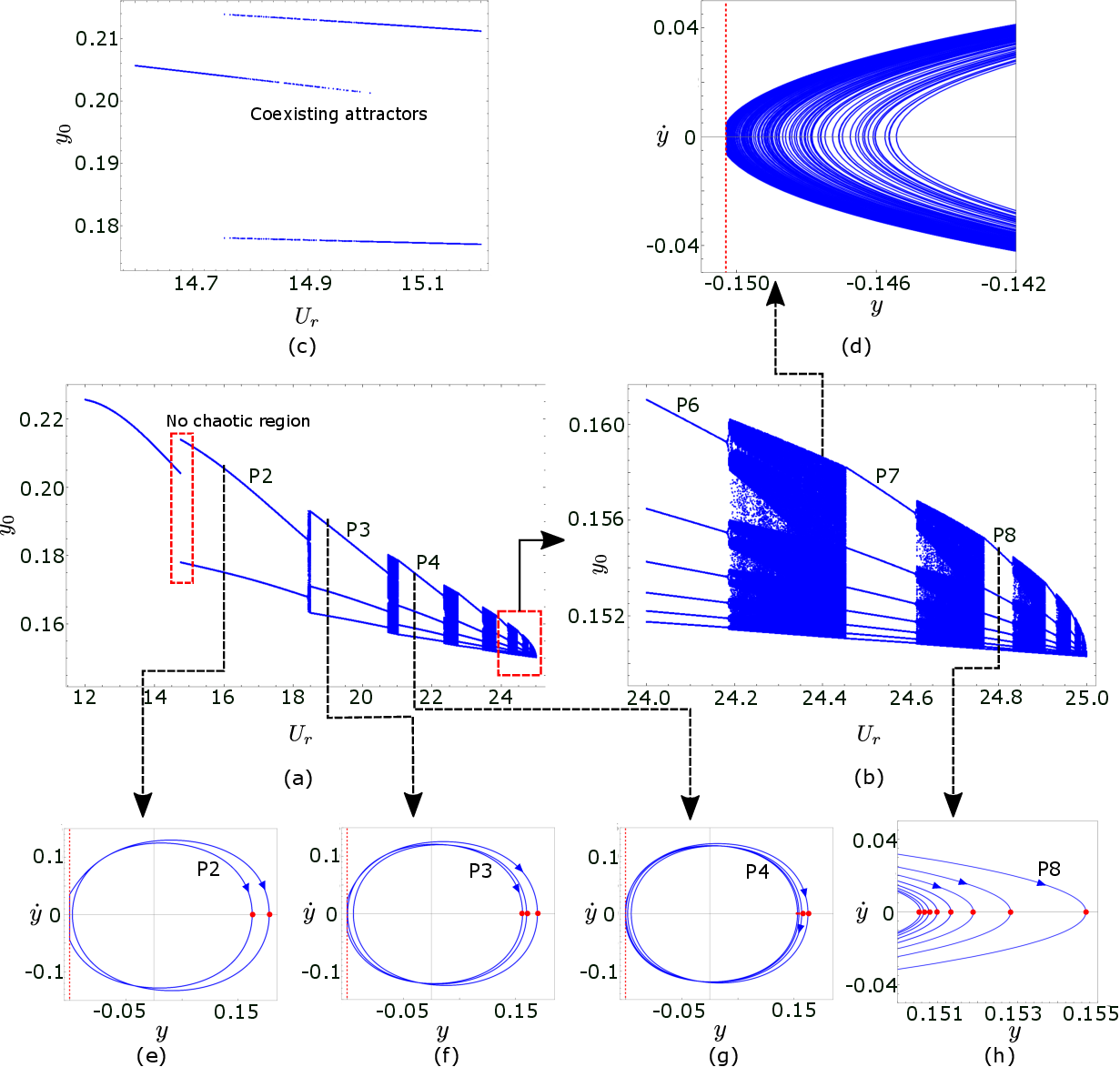}
    \caption{(a) Bifurcation diagram of the structure's amplitude response $y_0$ by varying reduced velocity $U_r$ for $r = 0.8$ and $\sigma = -0.150298$. (b) Amplitude response $y_0$ for $24.0 \leq U_r \leq 25.0$. (c) Bifurcation diagram depicting the coexistence of P1 and P2 attractors. (d) A typical aperiodic trajectory in phase-space for $U_r = 24.4$. Periodic orbits: (e) P2 when $U_r = 16.0$, (f) P3 when $U_r = 19.0$, (g) P4 when $U_r = 21.5$, (h) P8 when $U_r = 24.8$. Dynamics between $500$-$700$ impacts has been presented.}
    \label{fig 2}
\end{figure*}

In Fig. \ref{fig 2}, the amplitude response when the barrier is placed at $\sigma_{impact} = -0.150298$ with varying reduced velocity is shown. $\sigma_{impact}$ corresponds to the grazing boundary when $U_r = 25$. Note that an infinitesimal change in the critical $\sigma$ value may lead to a completely different orbit. Consequently, the value of $\sigma_{impact}$ is considered up to six significant digits. Fig. \ref{fig 2}(a) is the bifurcation diagram of structural displacement $y_0$ observed stroboscopically for varying reduced velocity ranging between $12 \leq U_r \leq 25.0$. The figure reveals the existence of period-adding cascades of orbits interspersed with bands of seemingly aperiodic responses. A zoomed section for the range $24.0 \leq U_r \leq 25.0$ is presented in Fig. \ref{fig 2}(b) for clarity. Figs. \ref{fig 2}(e) - \ref{fig 2}(h) show P2, P3, P4 and P8 orbits for $U_r = 16$, $19$, $21.5$ and $24.8$ respectively. The red dots highlight Poincar\'e points on $\Sigma$,
which have been plotted in Fig. \ref{fig 2}(a) and Fig. \ref{fig 2}(b). In Fig. \ref{fig 2}(c), simulations from multiple initial states reveal the presence of coexisting P1 and P2 attractors. Although no aperiodic band is observed in this window, such bands appear and get subsequently wider with the addition of periods in the cascade. In Fig \ref{fig 2}(d) this aperiodic behaviour, as reflected in phase portrait, is shown for $U_r = 24.4$.

\begin{figure}[h!]
    \centering
    \includegraphics[scale = 0.3]{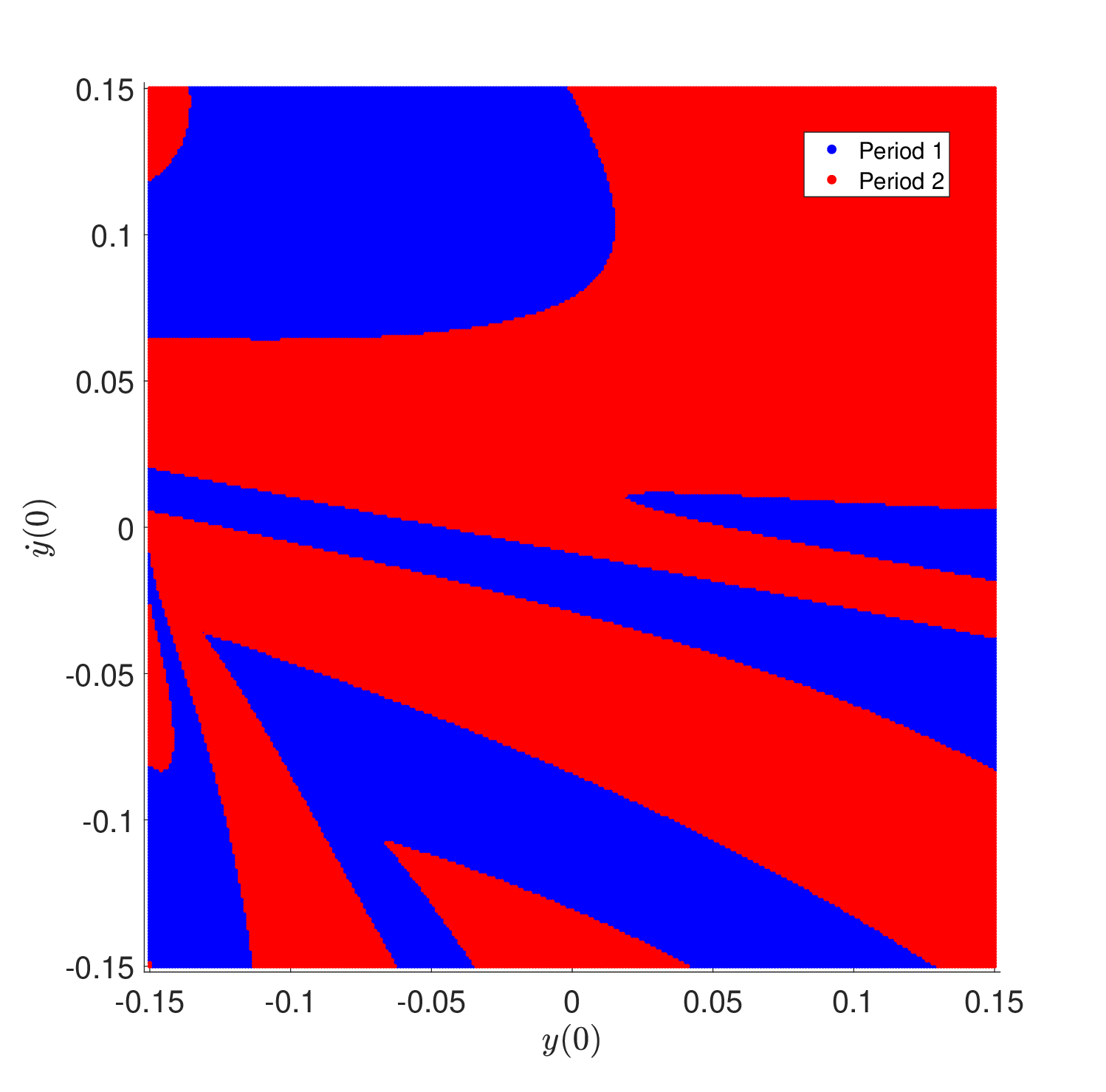}
    \caption{Basins of attraction for $U_r = 14.9$, $r = 0.80$ and $\sigma = -0.150298$.}
    \label{fig 3}
\end{figure}

\begin{figure}[h!]
    \centering
\includegraphics[scale = 0.30]{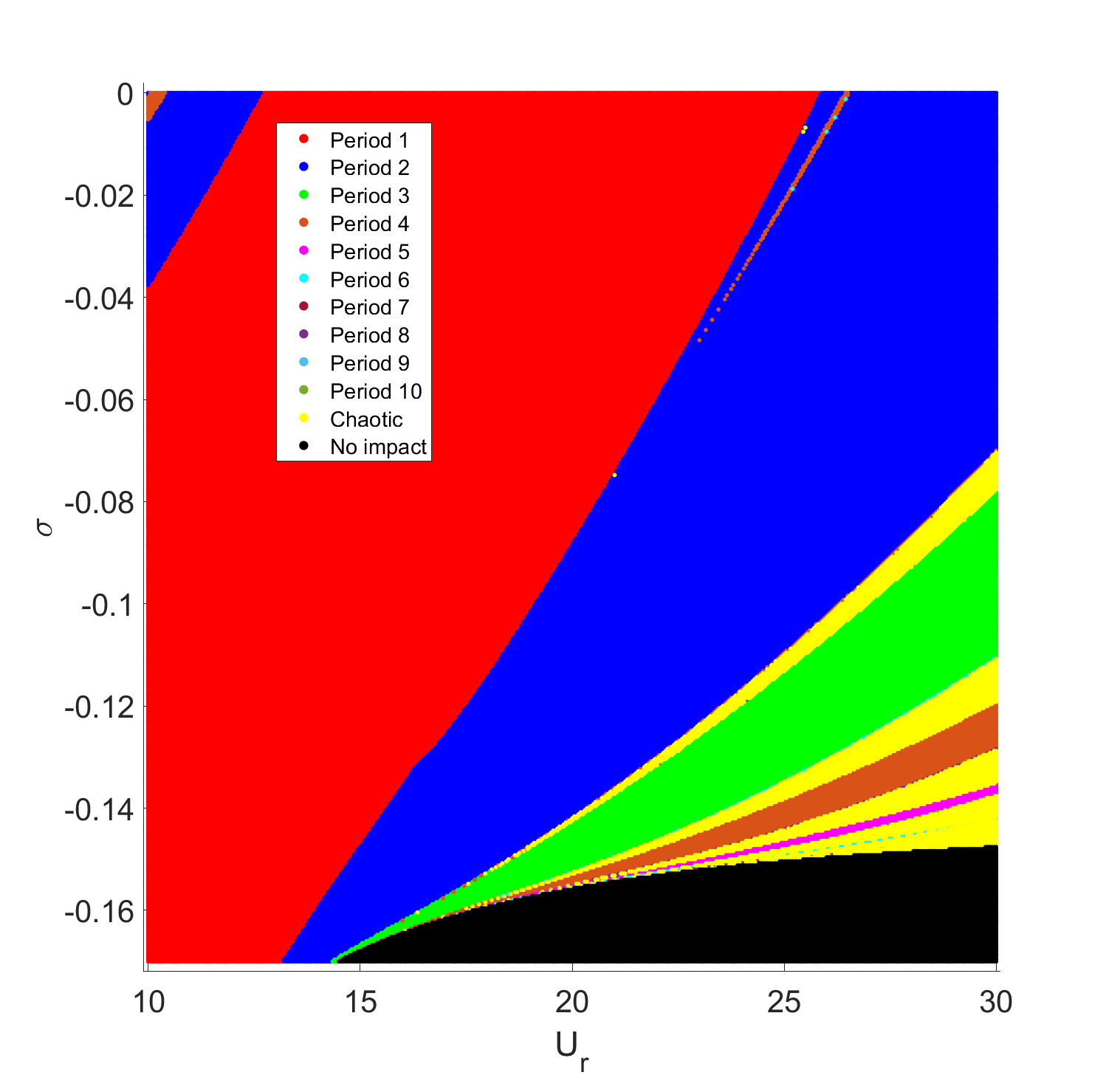}
    \caption{Periodicity in the dynamics of the structure's motion for $r = 0.8$. Periodicity has been calculated after $500$ impacts with a tolerance of $10^{-5}$.}
    \label{fig 4}
\end{figure}

The coexistence of P1 and P2 attractors in Fig. \ref{fig 2}(c) is further investigated by varying initial conditions while keeping $U_r$ and $\sigma$ fixed at $14.9$ and $-0.150298$, respectively. In Fig. \ref{fig 3}, the basins of attraction corresponding to P1 and P2 are shown. Results indicate that depending on the initial conditions, the steady-state orbits are either period one or two.  
Note that the placement of barrier at $\sigma_{impact} = -0.150298$ in Figs. \ref{fig 2} and \ref{fig 3} corresponds to the maximum amplitude of oscillations when $U_r = 25.0$. Thus, a barrier placed there would lead to grazing. A two-parameter bifurcation diagram is shown in Fig. \ref{fig 4}. 
The initial states $y(0)$, $\dot{y}(0)$, $x(0)$ and $\dot{x}(0)$ for these simulations were taken to be $0.0$, $0.0$, $0.0$ and $0.1$ respectively. The dynamics were observed on the Poincar\'e section $\Sigma$. 
The results are then colour-coded according to the periodicity for a given set of chosen parameters (i.e., $\sigma$ and $U_r$) post the transient cycles of $500$ impacts. Here, distinct regions of PN, aperiodic and no-impact {\it i.e,} P0 responses can be observed. Chaotic orbits were observed between various periodic orbits greater than one.

\begin{figure*}
    \centering
    \includegraphics[width=\linewidth]{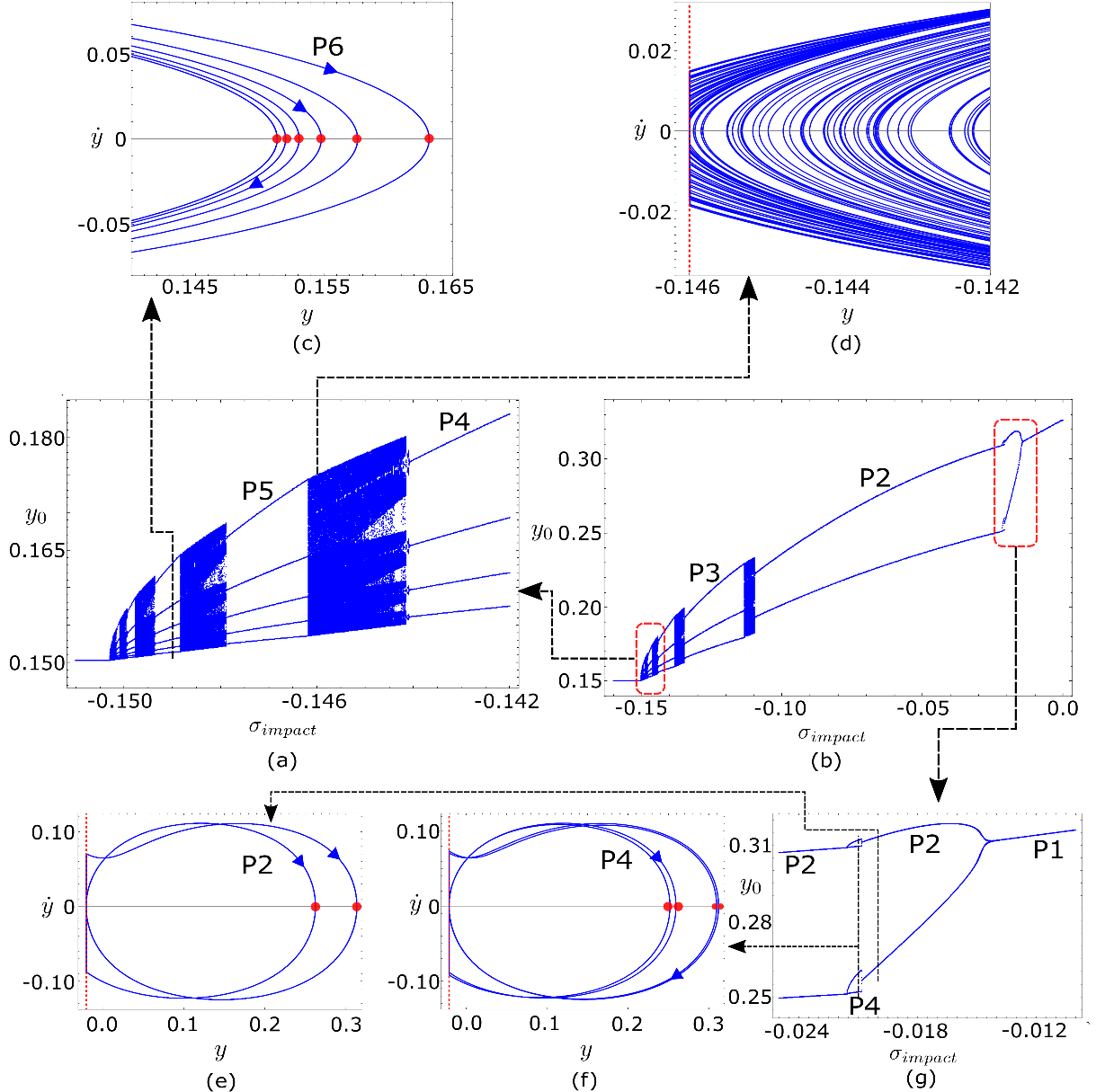}
    \caption{Bifurcation diagram depicting variation in structural amplitude $y_0$ as a function of $\sigma$ ranging between (a) $-0.150298 \leq \sigma \leq -0.142$ (b) $-0.150298 \leq \sigma \leq 0$ for $r = 0.8$ and $U_r = 25.0$. Phase portraits of: (c) P6 orbit when $\sigma = -0.149$, (d) Aperiodic orbit when $\sigma = -0.146$, (e) P2 orbit when $\sigma = -0.020$, (f) P4 orbit when $\sigma = -0.021$. (g) Bifurcation diagram $y_0$ when  $\sigma$ ranges between $-0.024 \leq \sigma \leq -0.010$. Dynamics between $500$-$700$ impacts are considered for these simulations.}
    \label{fig 5}
\end{figure*}

Next, the effect of barrier distance is investigated while keeping the fluid-flow velocity fixed. In Fig. \ref{fig 5}, barrier distance $\sigma$ is taken as the bifurcation parameter at a constant $U_r$ of 25. Fig. \ref{fig 5}(b) highlights the bifurcation behaviour when the oscillator is undergoing motion under pure compression {\it i.e.,} the barrier is placed such that the oscillator cannot reach its equilibrium position. Here, one can observe a period-adding cascade of orbits separated by bands of aperiodic solutions. In Fig. \ref{fig 5}(a), the bifurcation behaviour for $\sigma$ near the grazing condition (i.e., $\sigma_{impact})$ for $U_r = 25.0$ is shown. Higher period orbits are manifested when $\sigma \to \sigma_{impact}$. Fig. \ref{fig 5}(g) is a zoomed section of the bifurcation diagram where the amplitude response in the vicinity of P1 and P2 orbits can be observed near $\sigma = -0.020$. A bifurcation branch of a P4 orbit sandwiched between two P2 orbits can be observed here. This bifurcation curve branches out at an acute angle, signifying the occurrence of a non-smooth transition. In Figs. \ref{fig 5}(e) and \ref{fig 5}(f), the periodic orbits in phase-space corresponding to P2 and P4 for $\sigma = -0.020$ and $\sigma = -0.0210$, respectively, are shown. Similarly, the periodic orbit in phase-space corresponding to a P6 orbit is shown in Fig. \ref{fig 5}(c) for $\sigma = -0.149$. An aperiodic orbit in the phase-space, when the barrier is situated at $\sigma = -0.146$, is shown in Fig. \ref{fig 5}(d). These amplitude response diagrams show occurrences of bifurcations and the presence of seemingly chaotic orbits that were not observed in the absence of a rigid barrier \cite{govardhan2000modes}. Note that the bifurcation Figs. $2$ and $5$ show the Poincar\'e points obtained using direct numerical simulation and only correspond to the stable branches. Unstable branches cannot be determined using a direct numerical simulation method and require numerical continuation methods instead. This article does not present any bifurcation diagrams showing the unstable branches and will be taken up as a separate study in the future.   

\begin{figure*}
    \centering
    \includegraphics[width=\linewidth]{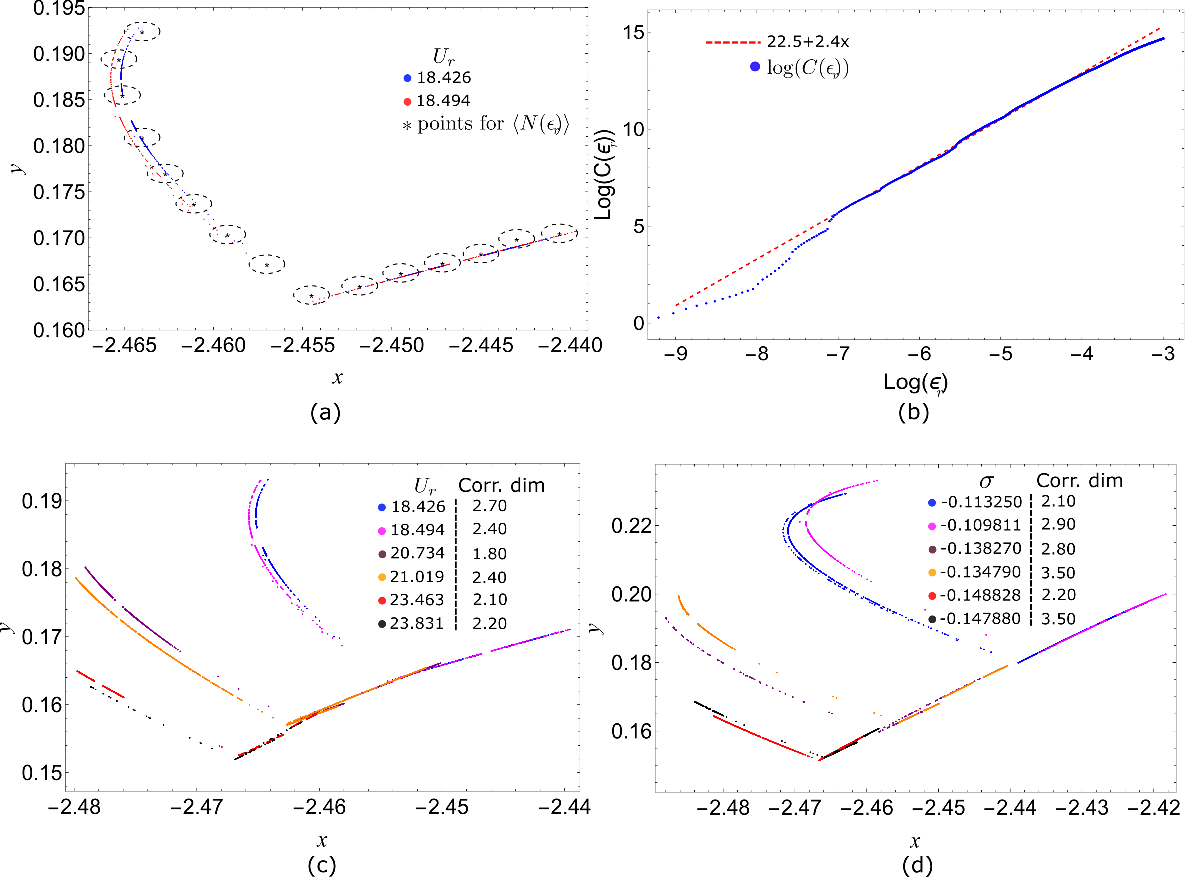}
    \caption{(a) Poincare section $\dot{y}=0$ for $U_r = 18.426$ and $18.494$. $N(\epsilon_r)$ is counted from the centre of circles shown; $C(\epsilon_r)$ defined as $\langle N(\epsilon_r)\rangle$. (b) Correlation dimension estimation from slope of $\log(C(\epsilon_r))$ vs $\log(\epsilon_r)$. Poincare section $\dot{y} = 0$ for (c) varying velocity $U_r$ and barrier kept fixed at $\sigma = -0.150298$; (d) varying barrier distance $\sigma$ with velocity fixed at $Ur = 25.0$.}
    \label{fig 6}
\end{figure*}

The aperiodic behaviour observed between the bifurcation branches of periodic orbits observed in Figs. \ref{fig 2} and \ref{fig 5} is investigated further. To establish that these aperiodic orbits are indeed chaotic, the dynamics on the Poincar\'e section $\Sigma$, governed by $\dot{y} = 0$, is examined. 
In Fig. \ref{fig 6}(a), trajectories intersecting $\Sigma$ for $U_r = 18.426$ and $18.494$ with a barrier located at $\sigma_{impact}$ and $r = 0.8$ are shown. To measure the dimension of the obtained Poincar\'e section \cite{grassberger1983measuring}, the correlation dimension $\nu$ has been calculated. This is done by counting the number of points $N(\epsilon_r)$ enclosed by a small circle of radius $\epsilon_r$ and letting the radius grow in size until it encloses the entire attractor. Since the points are clustered in different segments, multiple circles centred at different locations within the attractor are taken, and the mean of dimensions at these locations is considered as the entity $C(\epsilon_r)$ such that $C(\epsilon_r) = \langle N(\epsilon_r) \rangle$. The correlation dimension can then be estimated using the expression in Eq. \eqref{eq 5}
\begin{equation} \label{eq 5}
    C(\epsilon_r) \propto \epsilon_r^\nu.
\end{equation}

\noindent The number of centres for evaluation of $\nu$ are chosen as multiples of 2 {i.e.} $2$, $4$, $6 \ldots 12$. The values of $\nu$ tend to converge for centres more than $10$. In Fig. \ref{fig 6}(a), the mean of $N(\epsilon_r)$ is estimated by considering $15$ different centres that are circled in the figure. $\epsilon_r$ is varied from $0.0001$ to $0.06$ in steps of $10^{-4}$ to cover the entire finger-shaped attractor. The correlation dimension $\nu$ is then estimated from the slope of the log-log plot between $\epsilon_r$ and $C(\epsilon_r)$. A linear fit corresponding to the obtained graph is presented in Fig. \ref{fig 6}(b). Here, $\nu$ is estimated to be $2.4$. Since $\nu$ is a non-integer value, the corresponding dynamics observed on the Poincar\'e section possess a fractal-like structure and, hence, can be characterized as a strange attractor, indicative of the chaotic behaviour. In Fig. \ref{fig 6}(c), Poincar\'e sections are shown for various reduced velocities and barriers at $\sigma_{impact}$. The corresponding correlation dimensions obtained have non-integer values as well. In Fig. \ref{fig 6}(d) $\Sigma$ corresponding to varying barrier distances with reduced velocity kept fixed at $U_r = 25.0$ have been presented. The values of $U_r$ and $\sigma$ chosen here correspond to the aperiodic regions in the bifurcation diagrams that separate the periodic orbits {\it i.e.,} between P2-P3, P3-P4 and P5-P6 orbits. For each of these Poincar\'e sections, the correlation dimensions obtained are non-integer values, thus verifying the existence of fractal-like strange attractors sandwiched between periodic solutions.   

\begin{figure}
    \centering
    \includegraphics[scale = 0.29]{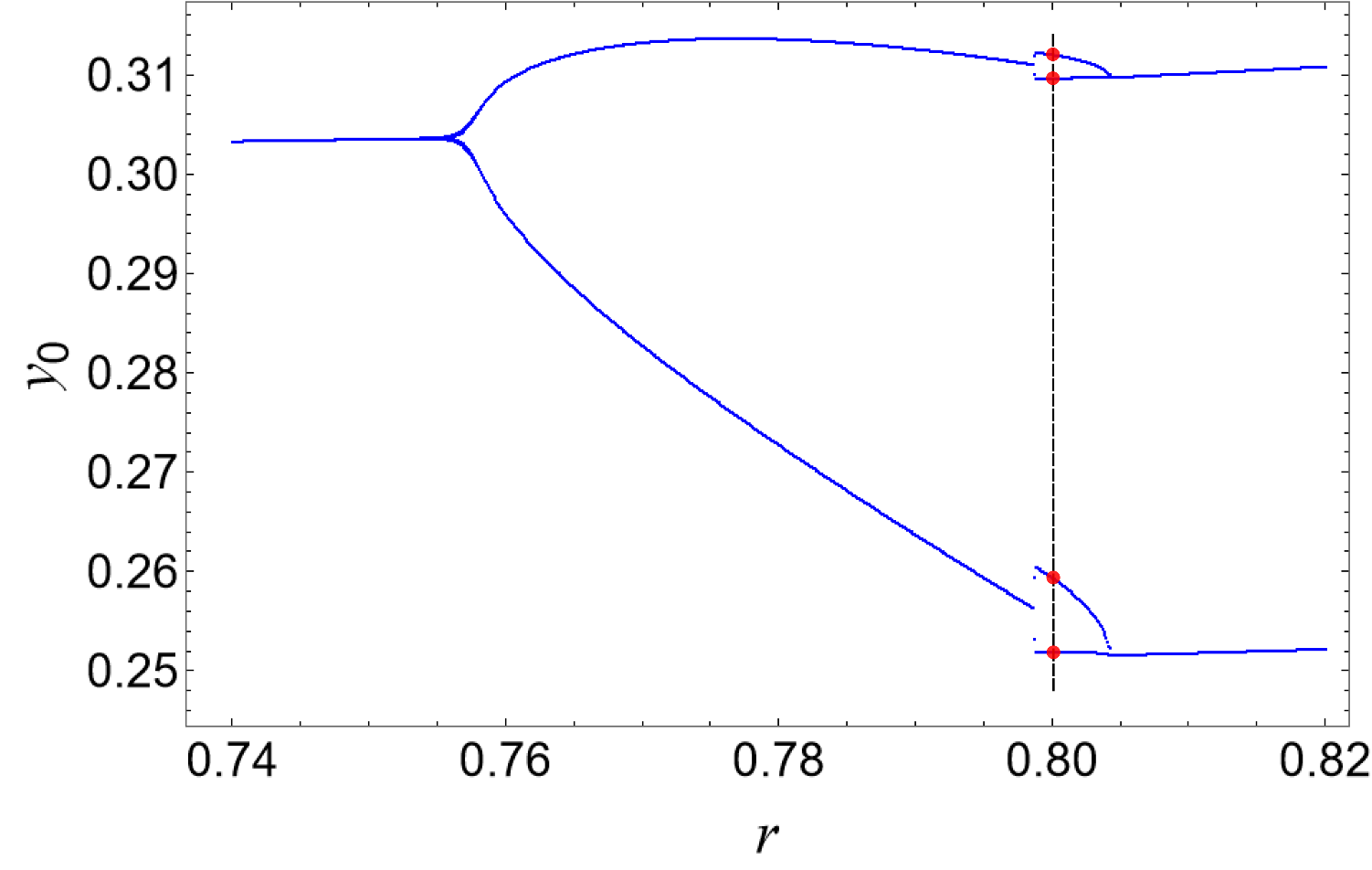}
    \caption{Bifurcation in the amplitude of the structure for various values of coefficient of restitution. Here, reduced velocity is fixed at $U_r = 25.0$ and with a barrier placed at $\sigma = -0.021$. The red dots correspond to Poincar\'e points for P4 shown in Fig. \ref{fig 5}(f).}
    \label{fig 7}
\end{figure}

\begin{figure}
    \centering
    \includegraphics[scale = 0.30]{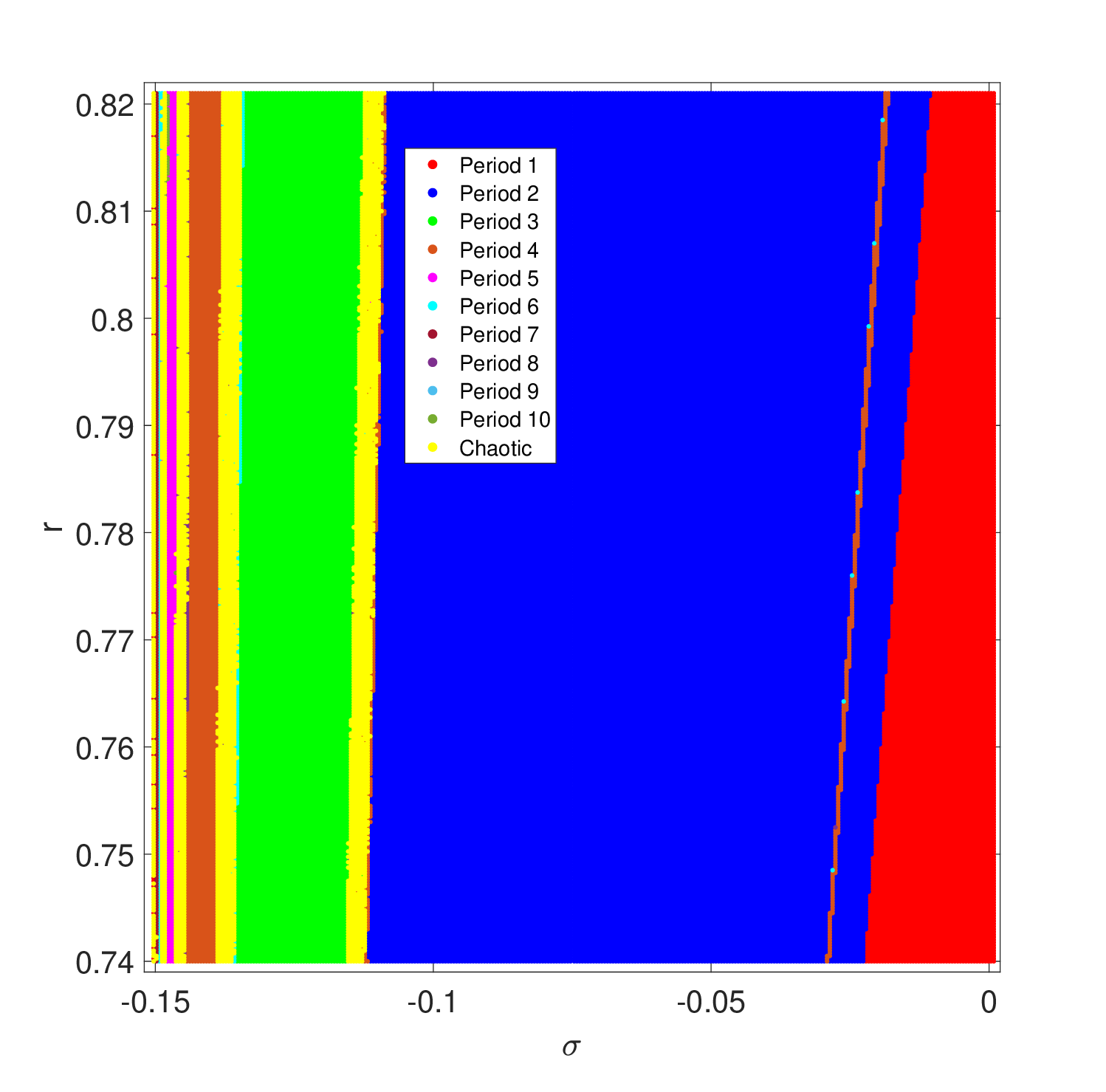}
    \caption{Two-parameter bifurcation diagram plotted between $\sigma$ and $r$ for reduced velocity $U_r = 25.0$.}
    \label{fig 8}
\end{figure}

\begin{figure}
    \centering
    \includegraphics[scale = 0.30]{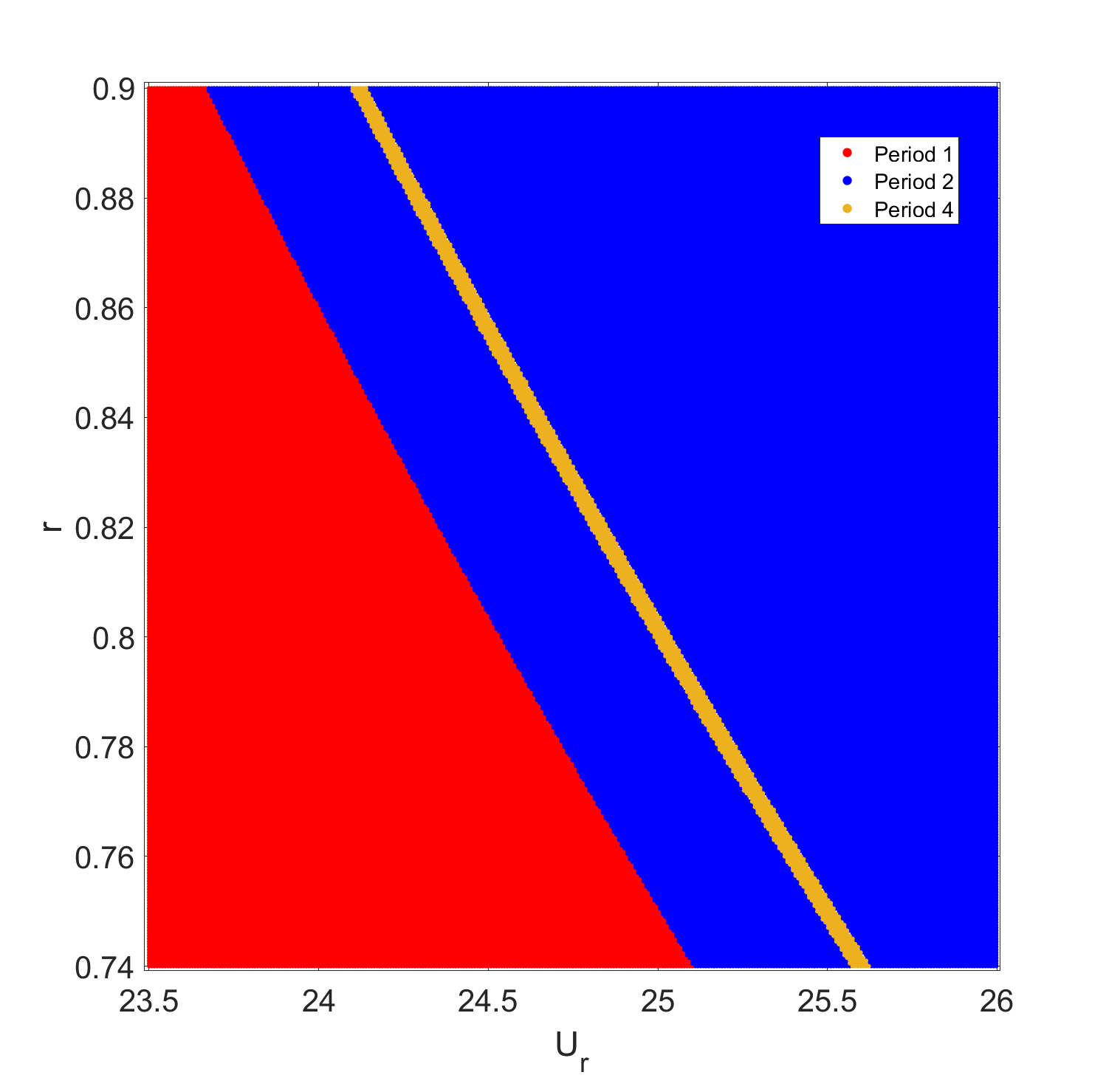}
    \caption{Two-parameter bifurcation diagram between $U_r$ and $r$ plotted for $\sigma = -0.0210$.}
    \label{fig 9}
\end{figure}

The effect on the dynamics due to the variation of the coefficient of restitution $r$ is investigated next. In Fig. \ref{fig 7}, the variation in amplitude response is shown against the coefficient of restitution $r$ when the barrier is fixed at $\sigma = -0.021$ and reduced velocity fixed at $U_r = 25.0$. The red dots shown in Figs. \ref{fig 5}(f) and \ref{fig 7} correspond to the Poincar\'e points of a P4 orbit, when $U_r = 25.0$, $\sigma = -0.021$ and $r = 0.8$. The observed behaviour is similar to Fig. \ref{fig 5}(g), where a P4 orbit is sandwiched between two P2 orbits. The results indicate that bifurcations also occur as the coefficient of restitution is varied. In Fig \ref{fig 8}, a two-parameter bifurcation diagram is presented. Here, periodicity in dynamics is observed to follow a distinct adding pattern, separated by chaotic bands. The reduced velocity for these simulations is kept fixed at $U_r = 25.0$. 
In the observed period-adding cascade, higher period orbits tend to appear in the vicinity of the grazing region that corresponds to $\sigma = -0.150298$ for $U_r = 25.0$. A narrow region of P4 orbit is observed between two P2 regions; see Fig. \ref{fig 8}. Using $U_r$ as one of the bifurcation parameters, similar behaviour is observed. The stable behaviour here is examined by keeping the barrier fixed while varying the reduced velocity and coefficient of restitution. This is shown in Fig. \ref{fig 9}. The barrier is kept fixed at $\sigma = -0.0210$, away from the grazing condition for $U_r = 25.0$. This distance corresponds to a value where P4 orbit was observed in Fig. \ref{fig 8} with $r = 0.8$ and $U_r = 25.0$. These discontinuity-induced bifurcation diagrams provide transition boundaries of the present system from one dynamical state to another. These results can inform the design of necessary clearances between multiple structures exhibiting VIVs to prevent fatigue and failure due to impacts. These responses can be avoided using design features that avoid such parameter values.

\begin{figure*}
    \centering
    \includegraphics[scale = 0.79]{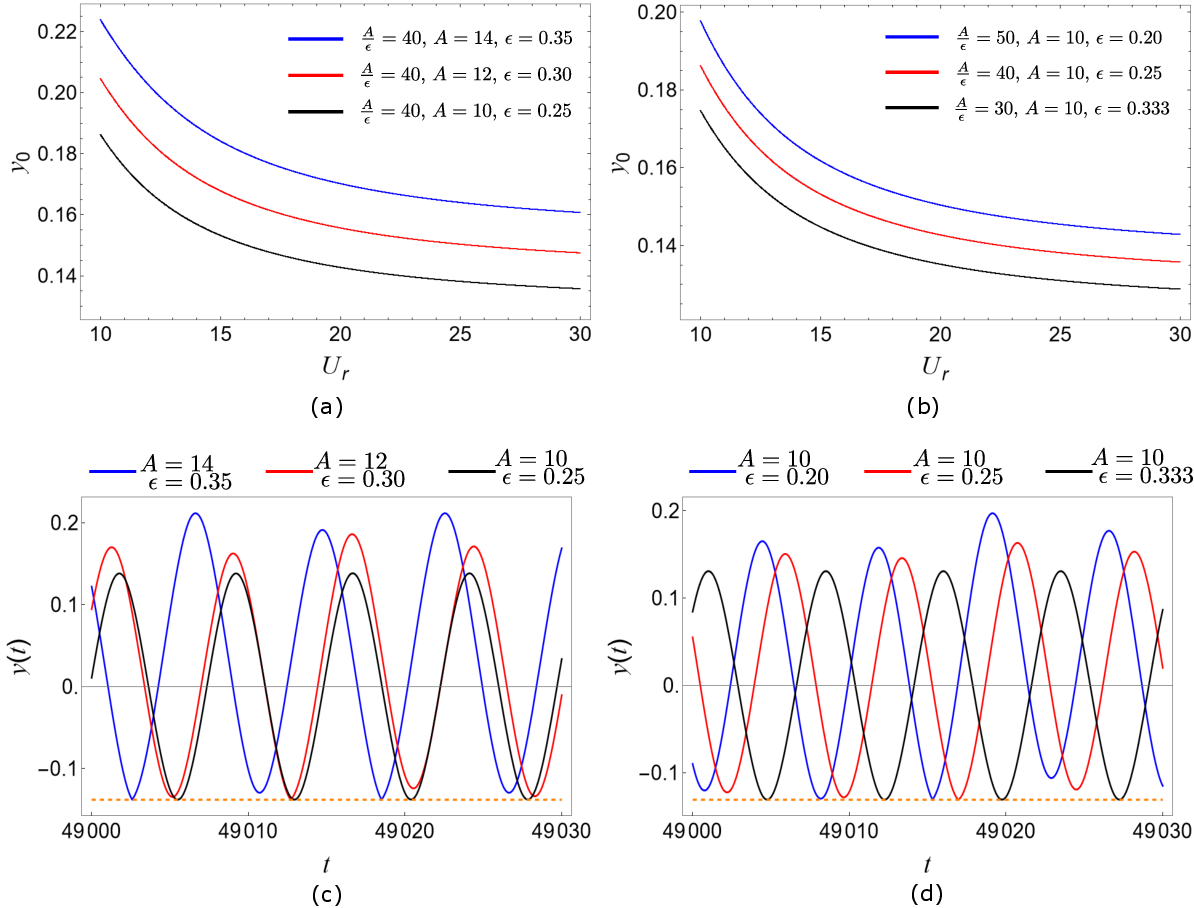}
    \caption{Amplitude response without barrier for (a) $A = 14$, $12$ and $10$ such that $A/\epsilon = 40$, (b) $A/\epsilon = 50$, $40$ and $30$. Structural displacement with barrier placed at (c) $\sigma = -0.13181$ by varying $A = 14$, $12$ and, $10$ ensuring $A/\epsilon = 40$. (d) $\sigma = -0.130979$ for $A/\epsilon = 50$, $40$ and $30$.}
    \label{fig 10}
\end{figure*}

For the above simulations, the coupling and the wake parameters, $A$ and $\epsilon$, are taken as $12$ and $0.3$, respectively, such that $A/\epsilon = 40$. This ensures that the dynamical model is compatible with experimental observations in \cite{facchinetti2004coupling}. The parameter $A$ determines the strength of coupling between the structure and associated wake dynamics, while $\epsilon$ controls the strength of the non-linearity of the wake oscillator. In Fig. \ref{fig 10}(a), the variation in amplitude response of the structure as a function of reduced velocity in the absence of a barrier is presented. The values of $A$ and $\epsilon$ are varied while ensuring that the ratio $A/\epsilon = 40$ is constant. It is observed that the amplitude of vibrations of the structure increases as the coupling parameter $A$ increases. In Fig. \ref{fig 10}(b), the amplitude response is shown in the absence of a barrier where $A$ is kept fixed at $A = 10$ and the ratio $A/\epsilon$ is varied. Note that the ratio of 40 ensured that the observed dynamics corresponded to experiments \cite{govardhan2000modes}; see the models discussed by Facchinetti {\it et. al} \cite{bishop1964lift,facchinetti2004coupling,govardhan2000modes}. In this case, the amplitude response increases as the ratio $A/\epsilon$ is increased. Next, the location of the rigid barrier is chosen such that, in the parameter regime of interest, the structure grazes the barrier for only one parameter. This implies that the vibration amplitude of the structure is the lowest in the domain of interest, ensuring that all other values of $A$ and $\epsilon$ will undergo an impact with the rigid barrier transversally. This is further demonstrated graphically. In Fig. \ref{fig 10}(c), the trajectories depicting the structure's oscillations are presented. The observations are made when the vibrations have attained a steady state for the barrier placed at $\sigma = -0.13181$. The barrier, shown as an orange dashed line, corresponds to the grazing condition for $A = 10$ and $\epsilon = 0.25$ such that $A/\epsilon = 40$. The black curve in Fig. \ref{fig 10}(c) grazes the barrier, while the blue and red curves interact with the barrier transversally. Similarly, in Fig. \ref{fig 10}(d), the structure oscillations are shown when the barrier is located at $\sigma = -0.130979$, corresponding to the grazing condition for $A = 10$, and $\epsilon = 0.333$, such that $A/\epsilon = 30$. Thus, the black curve with $A/\epsilon = 30$ grazes the barrier while the blue and red curves corresponding to $A/\epsilon = 50$ and $40$, respectively, impact the barrier transversally. A bifurcation analysis is subsequently carried out.

\begin{figure*}
    \centering
    \includegraphics[scale = 0.78]{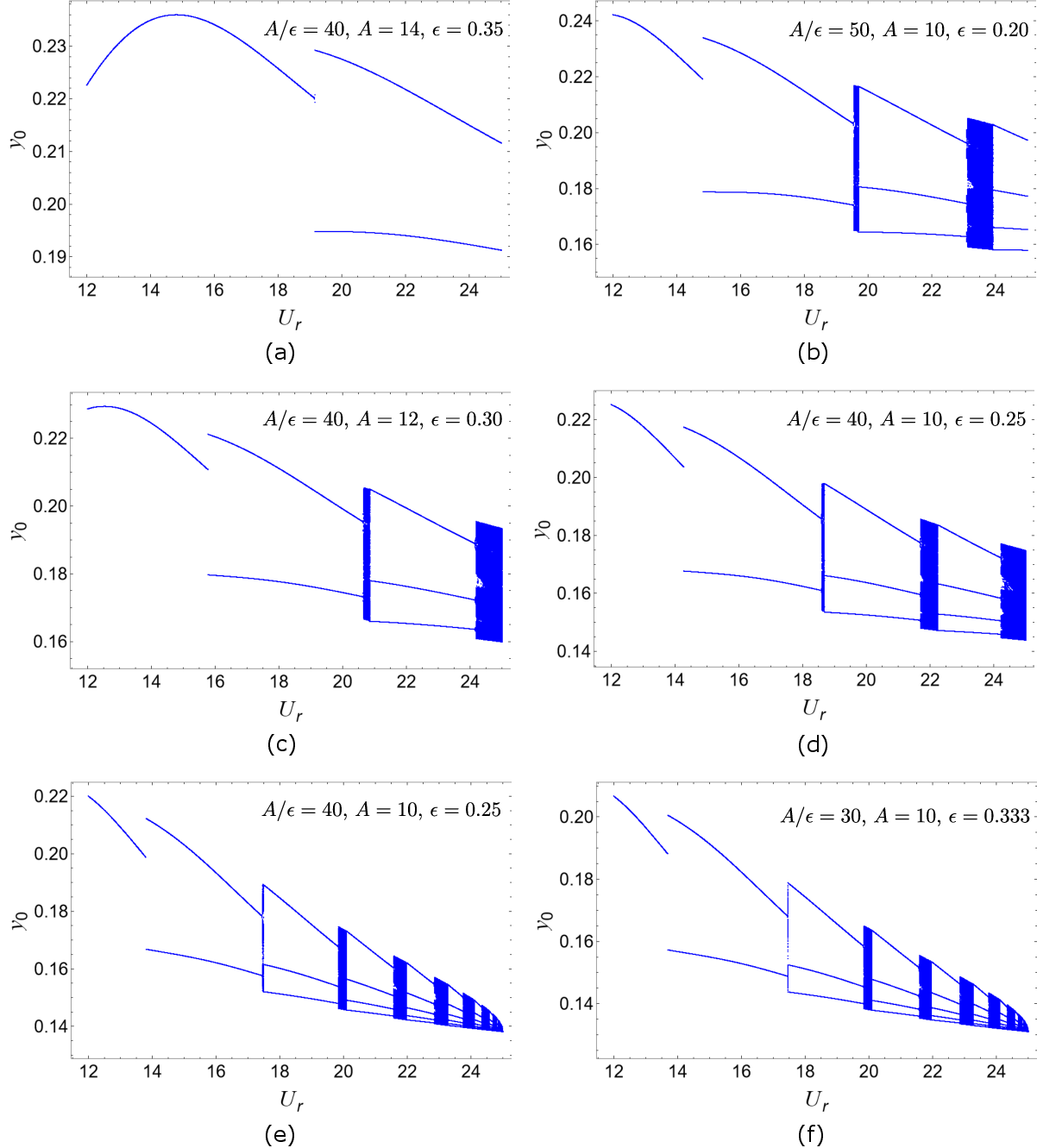}
    \caption{Bifurcation diagram with barrier placed at $\sigma = -0.13181$ for (a) $A = 14$ (c) $A = 12$ (e) $A = 10$ ensuring $A/\epsilon = 40$. Bifurcation diagram with barrier placed at $\sigma = -0.130979$ and $A = 10$ such that (b) $A/\epsilon = 50$, (d) $A/\epsilon = 40$ (f) $A/\epsilon = 30$.}
    \label{fig 11}
\end{figure*}

Figs. \ref{fig 11}(a), (c) and (e) depict the dynamical change in the amplitude response of the structure as a function of reduced velocity. The initial $500$ impacts are discarded to remove any transient effects and the ratio in all the cases is $A/\epsilon = 40$. The discontinuity boundary kept fixed at $\sigma = -0.13181$, which corresponds to the grazing condition for $A = 10$, $\epsilon = 0.25$ and $U_r = 25.0$. For $12 \leq U_r \leq 25.0$, only P1 and P2 orbits can be observed for $A = 14$ and $\epsilon = 0.35$; see Fig. \ref{fig 11}(a). As the coupling parameter $A$ reduces, the maximum amplitude of vibration of the structure reduces, thus leading to non-transversal interactions of trajectories with the boundary. As the value of $A$ approaches $10$ for which $\sigma$ is near grazing value of $-0.13181$, higher period orbits separated by chaotic bands are observed; see Figs. \ref{fig 11}(c) and (e). When $A = 10$ and $\epsilon = 0.25$ in Fig. \ref{fig 11}(e), the existence of higher periodic orbits is observed in the vicinity of $U_r = 25.0$ beyond which no impact with the barrier occurs. Similarly, in Figs. \ref{fig 11}(b), (d) and (f), the amplitude response of the cylindrical structure is shown for $A/\epsilon = 50$, $40$ and $30$, ensuring $A = 10$. The barrier is situated at $\sigma = -0.130979$, corresponding to the grazing condition for $A = 10$, $\epsilon = 0.333$ and $U_r = 25.0$. Again, a similar phenomenon is observed. With $A = 10$ and decreasing $A/\epsilon$, the amplitude of oscillations of the structure decreases, and the cylinder interacts with the barrier more non-transversally. A cascade of periodic orbits separated by chaotic behaviour is observed. As $A$, $\epsilon$ and $U_r$ approach the grazing condition of $\sigma = -0.130979$, higher periodic orbits are observed with bands of chaotic solutions; see Fig. \ref{fig 11}(f). However, in all of these results, no chaotic solutions are observed between P1 and P2 orbits. Results indicate that higher periodic orbits and chaotic trajectories are observed as the barrier is kept closer to the amplitude at which the cylinder grazes the rigid barrier. Such chaotic orbits can be avoided by ensuring enough clearance distance during the design of structures subjected to VIVs. 

The following section presents the stability analysis of this non-smooth system using Floquet multipliers and Lyapunov exponents. Here, the methods and numerical algorithms are presented, and inferences about the dynamical behaviour are drawn by correlating the results with the bifurcation studies presented above.

\section{Stability analysis} \label{sec 4}

Bifurcation diagrams presented above indicate that as parameters like $U_r$ or $\sigma$ is varied, existing orbits become unstable, resulting in newer periodic or aperiodic orbits. Such topological changes in the vibrational responses are known as discontinuity-induced bifurcations (DIBs) in the literature around non-smooth systems. DIBs can lead to unstable aperiodic solutions that are detrimental to the structure causing cause wear and tear. Hence, it is essential to carry out a stability analysis to investigate when an orbit might become unstable, therefore revealing routes to chaos.

Stability analysis of dynamical systems defined by piecewise-smooth vector fields is not straightforward. The stability of a periodic limit cycle for smooth dynamical systems can be predicted from the behaviour of perturbed trajectories in its local neighbourhood. The Jacobian of the vector field gives the dynamics of such perturbations. By measuring the rate of exponential divergence or convergence (Lyapunov exponents and Floquet multipliers) of perturbations from the primary limit cycle, the stability of the current orbit can be determined. However, for piecewise-smooth systems like the non-smooth FSI system above, the underlying dynamics change due to the presence of discontinuity barriers in the state space. Since perturbations to the system reach the discontinuity barrier at different time instances, the evaluation of the Jacobian matrices during an impact depends on how far the impact states of various perturbations are from the discontinuity barrier. This can be determined by the time it takes for various perturbations to reach the barrier. Therefore, a transverse discontinuity mapping (TDM) is necessary to ensure that perturbations reach the discontinuity barrier at the correct flight times. The TDM essentially stitches the two Jacobians calculated for perturbed trajectories during and after the instances of impacts. The geometrical interpretation and the derivation of the discontinuity mapping are discussed next.   

Consider an arbitrary steady-state dynamical system $\mathbf{x}$ with a small perturbation $\mathbf{y}$. The governing dynamics and its corresponding variational equations are presented in Eq. \eqref{eq 6}.
\begin{align} \label{eq 6}
    \dot{\mathbf{x}} &= \mathbf{F}(\mathbf{x}), \\
    \dot{\mathbf{y}} &\approx \mathbf{\nabla} \mathbf{F}(\mathbf{x})^T\cdot \mathbf{y} \nonumber.
\end{align}

\begin{figure}
    \centering
    \includegraphics[scale = 1.1]{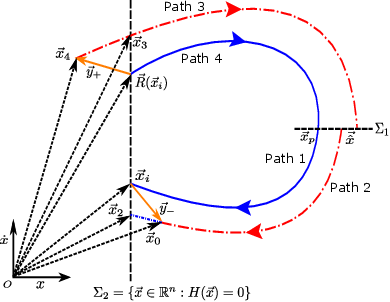}
    \caption{A schematic of the phase portrait of two nearby trajectories exhibiting impact at $\Sigma_{2}$. The blue line shows the actual trajectory. The red dashed line denotes the perturbed trajectory. This is a representative figure and is not to scale. The black dashed line $\Sigma_1$ denotes the Poincar\'e section from which both the trajectories are initiated and $\Sigma_2$ denotes the discontinuity boundary. The amber line denotes the perturbation vector $y$.}
    \label{fig 12}
\end{figure}

\noindent Fig. \ref{fig 12} is a schematic representation of two perturbed trajectories obeying Eq. \eqref{eq 6}. The periodic trajectory of the primary system starts initially from $\mathbf{x}_p$ and undergoes impact at $H(\mathbf{x}_i) = 0$. Here, $H(\mathbf{x}) = 0$ is a scalar function that models the discontinuity boundary.
However, any perturbed trajectory from this periodic orbit will not undergo an impact at the same instant when the primary orbit impacts the discontinuity boundary. Therefore, there exists a time difference between impacts for any two trajectories of a given hybrid system. The perturbed trajectory should be correctly mapped, accounting for the time difference until it impacts the barrier. Hence, the perturbation vector at the instant of impact is mapped from $\mathbf{y}_-$ to $\mathbf{y}_+$. This mapping ensures that the perturbed trajectory will reach the discontinuity boundary after the time difference between the impact of the two trajectories has elapsed. For a geometrical interpretation, let the primary state $\mathbf{x}_p$ impact $\Sigma_2$ at time $t_i$. At this instant, the perturbed path $\mathbf{\hat{x}}$ has reached $\mathbf{x}_0$ and further requires a time $\delta_1$ to reach $\Sigma_2$. At this instant, the primary state gets mapped from $\mathbf{x}_i$ to $\mathbf{R}(\mathbf{x}_i)$. However, trajectories in the local neighborhood, \textit{i.e.}, $\mathbf{x_0} = \mathbf{x}_i + \mathbf{y}_-$ has yet to reach the barrier. The transverse discontinuity mapping ensures that the state $\mathbf{x}_0$ and hence $\mathbf{y}_-$ reaches $\Sigma_2$ at $\mathbf{x}_2$ after time $t_i + \delta_1$, where $\delta_1$ can be approximated by a Taylor series expansion in the local neighbourhood of $\mathbf{x}_i$ \cite{bernardo2008piecewise, chawla2022stability}. After time $t_i + \delta_1$, the perturbed state $\mathbf{x}_2$ is mapped from $\Sigma_2 \rightarrow \Sigma_2$ using the reset map $\mathbf{x}_3 = \mathbf{R}(\mathbf{x}_2)$  

This can be alternately achieved by mapping the perturbed state from $\mathbf{x}_0$ to $\mathbf{x}_4$ at the instant $t = t_i$. Therefore, the perturbed trajectory initiated from $\mathbf{\hat{x}} = \mathbf{x}_p + \mathbf{y}_0$ reaches $\mathbf{x}_0 = \mathbf{x}_i + \mathbf{y}_-$ at the instant of impact of the primary trajectory and gets mapped from $\mathbf{y}_-$ to $\mathbf{y}_+$ such that $\mathbf{x}_4 = \mathbf{R}(\mathbf{x}_i) + \mathbf{y}_+$. Here, $\mathbf{R}(\mathbf{x}_i)$ represents the mapping of the primary periodic orbit (\textit{i.e.}, the impact of the structure with the barrier). The trajectory evolves from $\mathbf{x}_4$ and reaches $\mathbf{x}_3$ lying on the discontinuity boundary after a time $t_i + \delta_1$. This evolution can be interpreted as the perturbed trajectory $\mathbf{x}_0 = \mathbf{x}_i + \mathbf{y}_-$ undergoing impact at $H(\mathbf{x}_2) = 0$ after the time difference $t_i + \delta_1$ and getting mapped to $\mathbf{x}_3 = \mathbf{R}(\mathbf{x}_2)$. The above mapping of the perturbed trajectory from $\mathbf{y}_-$ to $\mathbf{y}_+$ can be expressed as a matrix transformation. This matrix transformation or simply a state transition matrix is known as a saltation matrix \cite{bernardo2008piecewise, muller1995calculation}; see Eq. \eqref{eq 7}.
\begin{equation} \label{eq 7}
    \mathbf{y}_+ = \mathbf{S} \cdot {\mathbf{y}_-}.
\end{equation}
The flight time $\delta_1$ and the saltation matrix $\mathbf{S}$ obtained by retaining only the first order terms of the Taylor series \cite{muller1995calculation,chawla2022stability} is presented in Eq. \eqref{eq 8}.
\begin{align} \label{eq 8}
    \delta_1 = &-\frac{\nabla H(\mathbf{x}_i) \cdot \mathbf{y}_-}{\nabla H(\mathbf{x}_i) \cdot \mathbf{F}(\mathbf{x}_i)}\\
   \mathbf{S} = &\mathbf{\nabla}\mathbf{R}(\mathbf{x}_i)^T \nonumber \\ &+ 
    \frac{\Big( \mathbf{F}(\mathbf{R}(\mathbf{x}_i)) - \mathbf{\nabla}\mathbf{R}(\mathbf{x}_i)^T\cdot\mathbf{F}(\mathbf{x}_i) \Big)}{\mathbf{\nabla}H(\mathbf{x}_i)^T\cdot\mathbf{F}(\mathbf{x}_i)} \otimes \mathbf{\nabla}H(\mathbf{x}_i)^T \nonumber
\end{align}
\noindent where $\mathbf{x}_i$ represents the state of the dynamical system at the instant of impact, $\mathbf{R}(\mathbf{x}_i)$ is the reset map at the instant of impact, $\mathbf{F}(\mathbf{x})$ denotes the underlying vector field and $H(\mathbf{x}) = 0$ represents the discontinuity boundary where an impact or reset occurs.

The discontinuity mapping and saltation matrix resolves the issue of discontinuities in the local neighbourhood of states subjected to impacts with a discontinuity barrier. Hence, conventional methods of stability analysis used for smooth dynamical systems can now be directly applied by incorporating the saltation matrix. The next section discusses two methods of investigating the stability of a limit cycle - an eigenvalue analysis (Floquet multipliers) of the monodromy matrix of a periodic orbit and an evaluation of its Lyapunov exponents.

\subsection{Floquet multipliers}

\begin{figure}
    \centering
    \includegraphics[scale = 0.9]{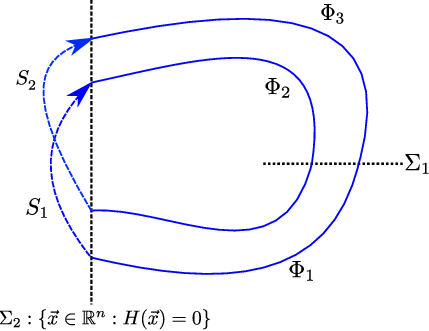}
    \caption{The schematic of a P$2$ limit cycle undergoing impacts twice with the discontinuity boundary at $\Sigma_2$, depicted as a black dashed line. The blue solid line corresponds to flow governed by the STMs $\mathbf{\Phi}_1$ to $\mathbf{\Phi}_3$ away from $\Sigma_2$ and the blue dashed line corresponds to state transition during impacts with $\Sigma_2$ (using saltation matrices $\mathbf{S}_1$ and $\mathbf{S}_2$).}
    \label{fig 13}
\end{figure}

The stability of orbits obtained for various parameters of the FSI system undergoing impacts is investigated next using Floquet multipliers. Since the system under parametric investigation is piecewise-smooth and comprises steady states separated by aperiodic orbits, a combination of Floquet theory along with the implementation of transverse discontinuity mappings are taken into consideration for calculating monodromy matrices. For smooth systems, Floquet multipliers are the eigenvalues of the global state transition matrix (STM) for any given periodic orbit. However, in hybrid dynamical systems, there are discontinuous transitions in the phase-space at the instant of an impact. Floquet multipliers are excellent indicators \cite{nayfeh2008applied} of when dynamical systems become unstable or will undergo a bifurcation. They can either change discontinuously, indicating the birth of a new orbit (saddle-node bifurcation), or can increase gradually, as some parameter is varied, until their magnitudes become greater than unity, thus indicating instability. For PWS dynamical systems, the global STM or the monodromy matrix is composed of local STMs and STMs at the instant of impact, also known as the saltation matrices. These matrices are multiplied in the order of occurrence of dynamical evolution. The state transition of any perturbed trajectory at the instant of an interaction with the barrier is defined by the TDM. The approach for obtaining the complete monodromy matrix and the role of appropriately defining the TDM is demonstrated next.  In Fig. \ref{fig 13}, a schematic phase portrait of a P2 orbit of a hybrid system undergoing impact twice with a discontinuity boundary $\Sigma_2$, depicted by a black dashed line, is shown. The monodromy matrix $\mathbf{\Phi}(T)$ for this orbit of period $T$ (time after which the orbit repeats itself) comprises products of several STM applied in the correct order of events. Here, the orbit is initiated on the section $\Sigma_1$, and it traverses to the switching surface $\Sigma_2$ by state transition $\mathbf{\Phi}_1$. The resultant state gets mapped by the saltation matrix $\mathbf{S}_1$ during impact. The mapping of this state back to $\Sigma_2$ again is given by $\mathbf{\Phi}_2$. It undergoes a state transition using the saltation matrix $\mathbf{S}_2$ and $\mathbf{\Phi}_3$ subsequently to land up at $\Sigma_1$, and the orbit repeats itself. Thus, over one cycle, the state transition matrix can be defined as
\begin{equation} \label{eq 9}
    \mathbf{\Phi}(T) = \mathbf{\Phi}_3 \cdot \mathbf{S}_2 \cdot \mathbf{\Phi}_2 \cdot \mathbf{S}_1 \cdot \mathbf{\Phi}_1,  
\end{equation}
where $\mathbf{S}_i$ is the STM or saltation matrix at the instant of $i^{th}$ impact. Similarly, for a given solution obtained numerically with period $T$ and $n$ number of impacts, the monodromy matrix is defined by Eq. \eqref{eq 10}.
\begin{equation} \label{eq 10}
    \mathbf{\Phi}(T) = \mathbf{\Phi}_n \prod_{i = 1}^{n-1} \mathbf{S}_i \cdot \mathbf{\Phi}_i. 
\end{equation}
According to the Floquet theory, the monodromy matrix is evaluated by integrating the fundamental solution matrix over one complete orbit. This ensures that the time-varying Jacobian over the limit cycle is now observed over one period, making the entries of the state transition matrix constant. Here, the saltation matrix describes the mapping of orbits in the local neighbourhood of the barrier at the instants when the dynamical system undergoes an impact with a discontinuity boundary. 

For the given FSI system described by Eq \eqref{eq 3} along with Eq \eqref{eq 4}, the state space is 4 dimensional {\it i.e.,} $\mathbf{x} = [y \ \dot{y} \ x \ \dot{x}]$ and the dynamics evolves according to the vector field $\mathbf{F}(\mathbf{x})$ described by Eq. \eqref{eq 11}.
\begin{equation} \label{eq 11}
    \mathbf{F} = \begin{bmatrix}
                    \dot{y}\\
                    -\delta^2 y - \alpha \dot{y} + M x\\
                    \dot{q}\\
                    -x - \epsilon(x^2 - 1)\dot{x}  + A(-\delta^2 y - \alpha \dot{y} + M x)\\
              \end{bmatrix},
\end{equation}
\noindent where $\alpha = (2\xi\delta + \gamma/\mu)$. The discontinuity barrier is defined using a scalar function $H(\mathbf{x}) = 0$ where $H(\mathbf{x}) = y - \sigma$. At the instant of impact a reset map $\mathbf{R}(\mathbf{x})$ is initiated such that $\mathbf{R}(\mathbf{x}) = [y \ -r \dot{y} \ x \ \dot{x}]$; $r$ being the coefficient of restitution defined previously in Eq. \eqref{eq 4}. The analytical form of the saltation matrix given by Eq. \eqref{eq 8} is represented in Eq. \eqref{eq 12},
\begin{equation} \label{eq 12}
    \mathbf{S} = \begin{bmatrix}
            -r & 0 & 0 & 0\\
            \frac{(1+r)}{\dot{y}}(-\delta^2 y + M x) & -r & 0 & 0\\
            0 & 0 & 1& 0\\
            A \alpha (1+r) & 0 & 0 & 1
        \end{bmatrix}.
\end{equation}

The fundamental solution matrix $\mathbf{Y}(t)$ is numerically evaluated by considering an orthogonal set of perturbed vectors such that $\mathbf{Y}(0) = \mathbb{I}_{n\times n}$. Each perturbation vector is a column entry of $\mathbf{Y}$ and evolves using Eq. \eqref{eq 6}. Next, the periodicity of the FSI system is obtained numerically after discarding $500$ impacts with the barrier to eliminate any transient effects. The monodromy matrix and its eigenvalues are evaluated after one time period from Eq. \eqref{eq 10}. The numerical procedure has been outlined in Algorithm \ref{algo 1}.
\begin{algorithm}[!]
    \caption{Floquet multipliers from monodromy matrix $\mathbf{\Phi}$} \label{algo 1}
    \begin{algorithmic}
    \State 1. Initialize: $\mathbf{x}(0)$, ensure $H(\mathbf{x}) > 0$
    \State 2. Initialize $U_r$ or $\sigma$ \Comment{Bifurcation parameter}
    \State 3. Initialize: $\mathbf{\Phi} = \textit{I}_{n \times n}$, count $= 0$
    \State 4. Initialize: $n_{max}$ \Comment{Maximum allowable impacts}
    \While{count $\leq$ $n_{max}$}
    \State Integrate: $\dot{\mathbf{x}} = \mathbf{F}(\mathbf{x})$
    \If{$H(\mathbf{x}) = 0$} \Comment{Occurrence of impact}
        \State Reset Map: $\mathbf{x} \gets \mathbf{R}(\mathbf{x})$
    \EndIf
    \If{$\dot{x} = 0$ $\&\&$ count $\geq n_{max}/2$}
        \State Store $\mathbf{x}$, $t$ for evaluation of time period $T$
        \State $T = \{t \in \mathbb{R}^1:\mathbf{x}(t + \tilde{n}T) = \mathbf{x}(t), \tilde{n}\in \mathbb{I} \}$
    \EndIf
    \If{count $=$ $n_{max}$} \Comment{Remove transients}
        \State Store: $\mathbf{x}_{init}$ $\gets \mathbf{x}(t)$ and $t_{init} \gets t$ 
    \EndIf
    \EndWhile
    \State Initialize: $\mathbf{x} \gets \mathbf{x}_{init}$ at $t_{init}$ \Comment{Begin at steady state}
    \State Initialize: $\mathbf{Y}(0) = \textit{I}_{n \times n}$ \Comment{Evaluation of $\mathbf{\Phi}_i$}
    \While{$t_{init} \leq t \leq (t_{init} + T)$} \Comment{Integrate over period $T$}
        \State Integrate: $\dot{\mathbf{x}} = \mathbf{F}(\mathbf{x})$
        \State Integrate: $\dot{\mathbf{Y}} = (\nabla \mathbf{F})^T\cdot\mathbf{Y}$
        \If{$H(\mathbf{x}) = 0$} \Comment{Occurrence of impact}
        \State $\mathbf{\Phi}_i \gets \mathbf{Y}$
        \State $\mathbf{\Phi} \gets \mathbf{\Phi}_i\cdot\mathbf{\Phi}$
        \State Evaluate $\mathbf{S}$ from Eq. \eqref{eq 12} \Comment{saltation matrix}
        \State $\mathbf{\Phi} \gets \mathbf{S} \cdot \mathbf{\Phi}$
        \State Reset Map: $\mathbf{x} \gets \mathbf{R}(\mathbf{x})$
        \State Reinitialize: $\mathbf{Y} \gets \textit{I}_{n \times n}$
        \EndIf
    \EndWhile
    \State 5. $\mathbf{\Phi} \gets \mathbf{Y}\cdot\mathbf{\Phi}$ \Comment{Evaluate monodromy matrix}
    \State 6. Evaluate eigenvalues of $\mathbf{\Phi}$ to get Floquet multipliers
    \end{algorithmic}
\end{algorithm}

\begin{figure*}
    \centering
    \includegraphics[scale = 0.4]{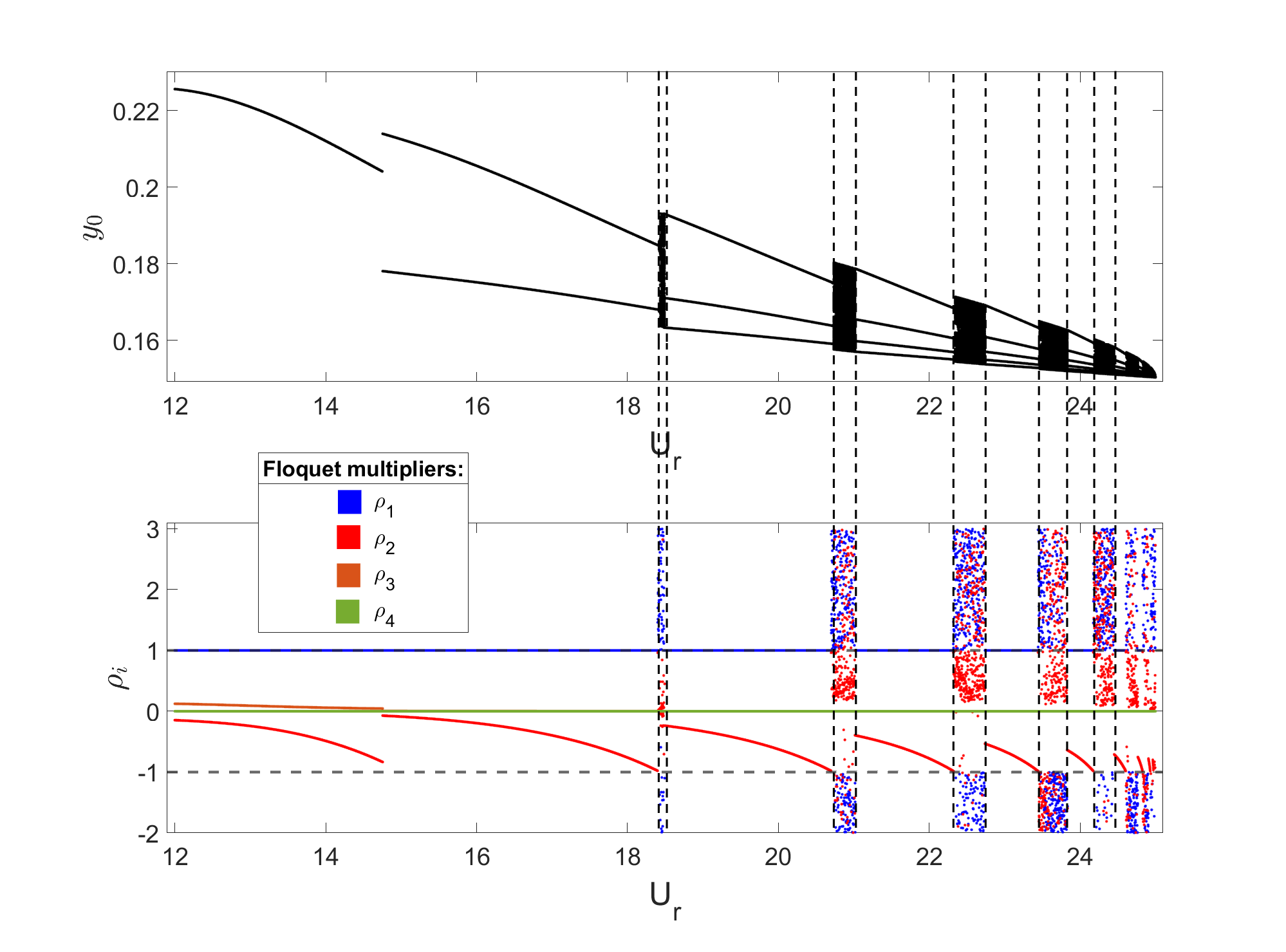}
    \caption{Floquet multipliers vs reduced velocity $U_r$ for barrier placed at $\sigma = -0.150298$ and $r = 0.8$.}
    \label{fig 14}
\end{figure*}

\begin{figure*}
    \centering
    \includegraphics[scale = 0.4]{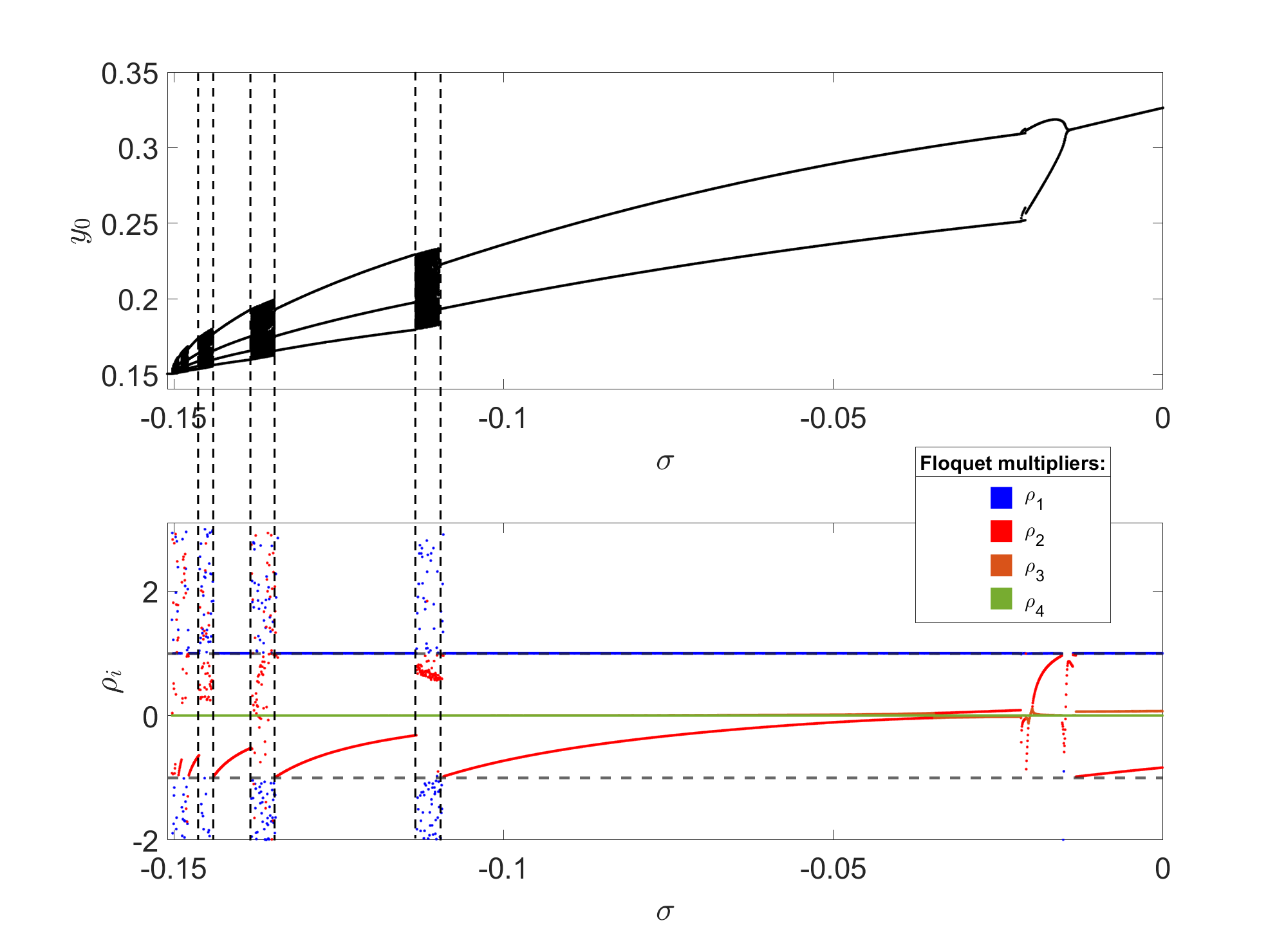}
    \caption{Floquet multipliers vs barrier distance $\sigma$ when reduced velocity is $U_r = 25.0$ and $r = 0.8$.}
    \label{fig 15}
\end{figure*}

In Fig. \ref{fig 14}, the Floquet multipliers $\rho_i$ of the monodromy matrix have been shown as the reduced velocity $U_r$ is varied. A one-to-one correspondence is drawn with the dynamics observed on the bifurcation diagram. A barrier is placed at $\sigma_{impact}$ (grazing condition for $U_r = 25.0$) with $r = 0.8$. The region ranging between $-1 \leq \rho_i \leq 1$ corresponds to a stable periodic orbit. The dashed vertical lines highlight the observed bands of chaotic orbits. Floquet multipliers are within the unit circle for stable orbits. As $U_r$ is varied, the Floquet multiplier approaches $-1$, eventually resulting in a bifurcation to chaotic orbits at $\rho_i = -1$. Additionally, in the parameter region where a transition from P1 to P2 orbit is observed, the Floquet multipliers discontinuously jump to a value with a norm less than unity. This suggests the birth of a new stable periodic orbit. Eventually, the absolute value of the Floquet multipliers decreases to $-1$ before it approaches the next instant of a non-smooth bifurcation. Similarly, in Fig. \ref{fig 15}, the Floquet multipliers are shown as the barrier distance is varied, while the reduced velocity is kept fixed at $U_r = 25.0$ and $r = 0.8$. Similar transitions in values of the Floquet multipliers are observed near bifurcation occurrences. This is clearly reflected in the bifurcation diagram as well.
Moreover, it was also observed that one of the Floquet multipliers is always unity, {\it i.e.} it lies on the unit circle. For an autonomous dynamical system, this implies that there is an eigendirection along that which there is no variation in the magnitude of perturbation {\it i.e.} the primary and the perturbed trajectory neither converge nor diverge from each other \cite{nayfeh2008applied}. This also verifies that the TDM correctly maps the perturbed trajectories at the instant of impact, and the results verify the established Floquet theory. It is to be noted here that Floquet's theory is not directly applicable to aperiodic oscillatory responses. This is because the trajectories do not recur, and thus, the product of matrix multiplications diverges. This is reflected in Fig. \ref{fig 14}(b) as noisy bands of $\rho_i$ with a magnitude greater than unity. An alternate marker of chaotic dynamics is the presence of positive Lyapunov exponents. Lyapunov spectra are thus obtained below to analyze the observed aperiodic phenomena in this non-smooth system.

\subsection{Lyapunov exponents}

In this section, the stability of an orbit is investigated by estimating the rate of exponential divergence or convergence of perturbed orbits in its local neighbourhood. For PWS dynamical systems, it is important to accurately know when perturbed orbits reach the discontinuity barrier and get mapped after an impact. This is given by the flight time estimate $\delta_1$ and the transverse discontinuity mapping of $\mathbf{y}_-$ to $\mathbf{y}_+$. A first-order approximation \cite{bernardo2008piecewise,muller1995calculation} of perturbed states after an impact \textit{i.e.}, $\mathbf{y_+}$ can become inaccurate after multiple impacts as shown in \cite{chawla2022stability}. A linearized approximation of the transverse discontinuity mapping \cite{bernardo2008piecewise} predicts that perturbations undergo impacts with the discontinuity barrier. However, discontinuity mappings, when derived using a higher-order Taylor series approximation, reveal that not all orbits in the local neighbourhood of the barrier should undergo impact - some orbits can miss the discontinuity barrier since they never reach it \cite{chawla2022stability}. This article implements a higher-order transverse discontinuity mapping \cite{chawla2022stability} that maps perturbations from $\mathbf{y}_-$ to $\mathbf{y}_+$ during an impact of the primary orbit at $\mathbf{x}_i$ with the discontinuity barrier. The rates of exponential divergence or the Lyapunov exponents are computed as follows. Starting from a periodic orbit $\mathbf{x}_0$, an orthogonal set of perturbed trajectories is initialized. The initial magnitude of the perturbed trajectories is set to a small value of $r_0 = 0.0001$ units. This implies that the perturbed vectors lie on a hypersphere of radius $r_0$. Then, this hypersphere is allowed to evolve for the $n^{th}$ time from $t$ to $t + t_0$, after which the growth rate $r_n$ of all the perturbed trajectories is noted. The perturbed trajectories evolve according to the variational formalism of the dynamical system under investigation (see Eq. \eqref{eq 6}). Each time the primary system undergoes an impact with the discontinuity boundary, the perturbed trajectory gets mapped by Eq. \eqref{eq a1} in phase-space using the higher-order TDM defined in Appendix A. After a time interval of $t_0$, the initialized hypersphere has now evolved into a hyper-ellipsoid. The average largest Lyapunov exponent $\lambda_1$ is measured from the growth rate of the perturbations along the major axis of the hyper-ellipsoid using Eq. \eqref{eq 13},
\begin{equation} \label{eq 13}
    \lambda_i = \frac{1}{t_0 N} \Sigma_{n = 1}^{N} \ln\frac{r_n}{r_0}.
\end{equation}
\noindent At this instant, the set of perturbed vectors is once again orthogonalized using the Gram-Schmidt process with respect to one of the perturbed vectors (along eigenvector with the largest Lyapunov exponent), meanwhile ensuring the magnitude is again scaled down to $r_0$. This is achieved by performing a QR decomposition (QRD) followed by scaling of the vectors by $r_0$. Thus, once again, an initial hypersphere of radius $r_0$ is obtained, but the vectors are now oriented differently. This process is repeated, and the growth rate is monitored and resets stroboscopically, {\it i.e.} when $t \% t_0 = 0$. This step is followed by a QRD. When a discontinuity boundary is encountered during an impact at $\mathbf{x}_i$, the perturbed orbit gets mapped from $\mathbf{y}_-$ to $\mathbf{y}_+$ using a transverse discontinuity mapping derived based on higher-order Taylor series approximations \cite{chawla2022stability} of the flow in the local neighbourhood of the barrier.
The higher-order transverse discontinuity mapping for the present non-smooth FSI system is given in Appendix A. This procedure has been described in Algorithm \ref{algo 2}. The Lyapunov exponents along each orthogonal direction are thus obtained by evaluating steps 6-8 in Algorithm \ref{algo 2}; also see Eq. \eqref{eq 13}.

\begin{algorithm}[h!]
    \caption{Lyapunov exponent for hybrid systems using $\mathcal{O}(1)$ TDM} \label{algo 2}
    \begin{algorithmic}
    \State 1. Initialize: $\mathbf{x}(0)$ , ensure $H(\mathbf{x} \geq 0)$
    \State 2. Initialize: $U_r$ or $\sigma$ \Comment{Bifurcation parameter}
    \State 3. Initialize: $\mathbf{y}_i(0)$ for $i \leq n$ using QRD
    \State 4. Rescale: $\mathbf{y}_i(0) \gets r_0*\mathbf{y}_i(0)$
    \State 5. Initialize: $n_{max}$ \Comment{Maximum allowable impacts}
    \While{count $\leq n_{max}$}
    \State Integrate: $\dot{\mathbf{x}} = \mathbf{F}(\mathbf{x})$
    \If{$H(\mathbf{x}) = 0$} \Comment{Occurrence of impact}
        \State Evaluate: $\mathbf{R}(\mathbf{x})$ and $\mathbf{y}_+$
        \State Reset Map: $\mathbf{x} \gets \mathbf{R}(\mathbf{x})$
        \State Reinitialize: $\mathbf{y}_i \gets \mathbf{y}_{i,+}$ \Comment{higher-order TDM defined in Eq. \eqref{eq a1} of Appendix A}
    \EndIf
    \If{$\big( t \% t_0 \big) = 0$}
        \State \If{count $\geq n_{max}/2$}
                    \State Store: $r_i \gets \frac{1}{r_0} \| \mathbf{y}_i \|$
               \EndIf
        \State Reinitialize: $\mathbf{y}_i \gets r_0 * \text{QRD of } \mathbf{y}_i$ 
    \EndIf
    \EndWhile
    \State 6: Evaluate: $\ln r_i$ \Comment{Store all $\ln r_i$}
    \State 7: LE$_i \gets \frac{1}{t_0 N} \Sigma_{n = 1}^N \ln \frac{r_n}{r_0}$ \Comment{LEs measured N times}
    \State 8: $\lambda_i \gets \langle \text{LE}_i \rangle$ \Comment{Mean of all LEs}
    \end{algorithmic}
\end{algorithm}

\begin{figure*}
    \centering
    \includegraphics[scale = 0.4]{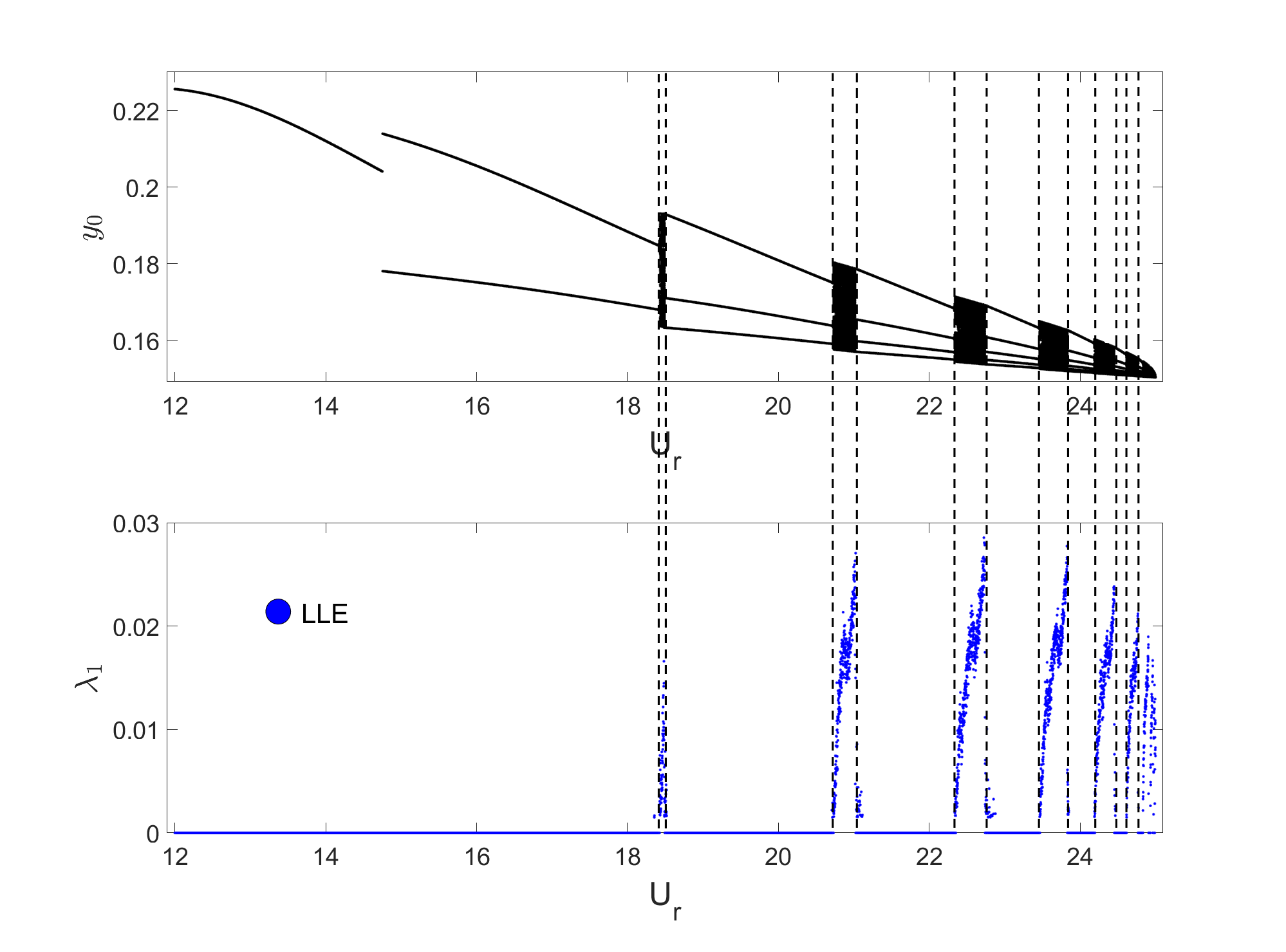}
    \caption{Largest lyapunov exponent vs reduced velocity $U_r$ when barrier is placed at $\sigma = -0.150298$ and $r = 0.8$.}
    \label{fig 16}
\end{figure*}

\begin{figure*}
    \centering
    \includegraphics[scale = 0.4]{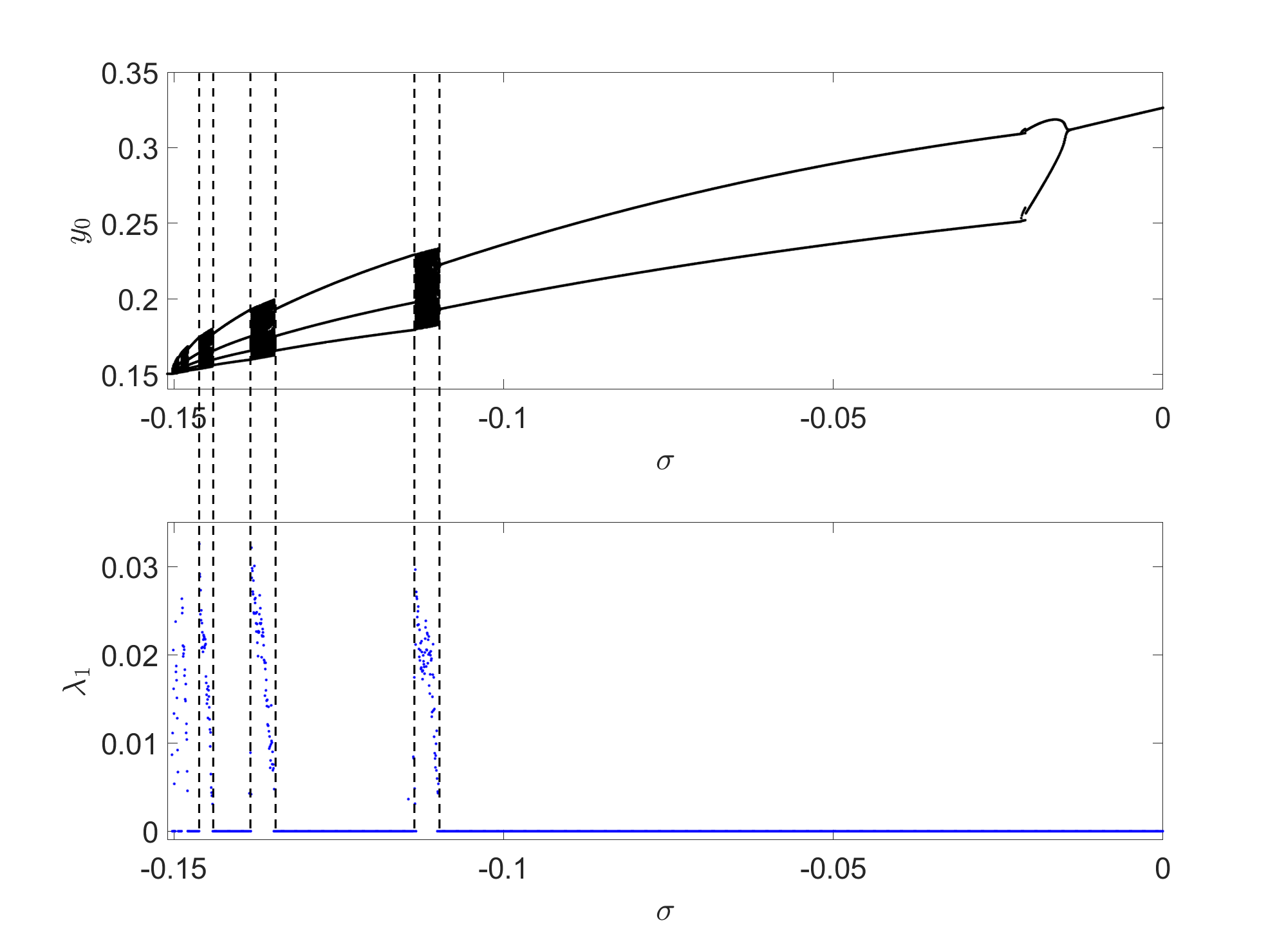}
    \caption{Largest lyapunov exponent vs barrier distance $\sigma$ when reduced velocity is $U_r = 25.0$ and $r = 0.8$.}
    \label{fig 17}
\end{figure*}

In Fig. \ref{fig 16}, the variation of the largest Lyapunov exponent (LLE) as a function of reduced velocity, as well as the corresponding bifurcation diagrams, are shown. The barrier is kept fixed at $\sigma_{impact}$ and $r = 0.8$. The LLE has been computed stroboscopically at $t_0 = \pi$ in algorithm \ref{algo 2} for $800$ impacts with the barrier. The LLE is a measure of the maximum rate of exponential divergence between two nearby perturbed trajectories. Positive values of LLE in Fig. \ref{fig 16} correspond to trajectories that diverge from each other, indicating the occurrence of chaos. This can also be verified from the corresponding bifurcation diagram. The dashed lines in the figure highlight the intervals within which chaotic orbits are observed. These regions also show a positive value of the LLE. Similarly, in Fig. \ref{fig 17}, LLEs have been shown as the barrier distance is varied, ranging between $-0.150298 \leq \sigma \leq 0$. The flow velocity here is kept fixed at $U_r = 25.0$ and $r = 0.8$. The LLE is zero for all periodic orbits and suddenly jumps to positive values where chaotic orbits are observed in the corresponding amplitude response diagram. 

Therefore, the higher-order TDM accurately determines the dynamical stability of the FSI system undergoing impacts. The algorithms presented above can also be adapted to consider higher-order estimations of the TDM for better estimation of the states near an impact \cite{chawla2022stability}. Furthermore, the proposed algorithms can be used to study non-smooth dynamical systems of different formalisms {\it i.e.}, the Filippov and piecewise continuous kind and can be extended to higher dimensions. The findings from the presented work are summarized in the section below.    
\section{Conclusions} \label{sec 5}

The effect of impacts with a rigid barrier on the phase-locked behaviour of a sdof cylindrical system is investigated in this paper. A detailed study of the underlying dynamical bifurcations, followed by the stability analysis of the concerned FSI system undergoing vibro-impacts, is presented. 
Non-smooth dynamical phenomena like period adding, periodic bifurcation branches separated by chaotic orbits, and finger-shaped chaotic attractors are observed. Moreover, P1 and P2 stable attractors with intertwined basins of attraction coexist. 
By varying the coupling strength and non-linearity of the wake oscillator, cascades of higher periodic orbits were observed in the vicinity of the grazing condition.

Furthermore, algorithms for obtaining Floquet multipliers and Lyapunov spectra have been modified to account for non-smoothness in dynamical systems. This is done by incorporating non-smooth saltations into the state transition matrices. The effectiveness of these proposed algorithms is demonstrated for the vibro-impact system under fluid flow. To capture the discontinuous resetting of the orbits in phase space due to the interaction with a non-smooth barrier, a transverse discontinuity map is analytically derived at the instant of impact. Higher-order correction terms are retained in the transverse discontinuity mapping, and the effectiveness of this approach is verified by the corresponding largest Lyapunov exponent. Further, numerically obtained Floquet multipliers have effectively predicted occurrences of discontinuity-induced bifurcations as they approach the critical value of unity as parameters vary gradually here, such as the barrier distance and reduced flow velocity. Thus, Floquet multipliers are an excellent choice for the investigation of autonomous piecewise-smooth dynamical systems. Additionally, sudden jumps in the Floquet multipliers, indicating the occurrence of a non-smooth bifurcation, are observed when one of the orbits loses stability, and another comes into existence upon variation of a system parameter. The parameter corresponding to the loss of system stability, as observed with the bifurcation diagrams, corresponds exactly with the parameter when the Floquet multipliers exit the unit circle and the largest Lyapunov exponent is positive, verifying the accuracy of the proposed approach. 

\section*{Declarations}
The authors acknowledge the support of the Sustainable Energy Authority of Ireland (SEAI) TwinFarm project RDD/604, Science Foundation Ireland NexSys Project 21/SPP/3756, Enterprise Ireland SEMPRE Project and SEAI FlowDyn project RDD/966.

\section*{Conflict of interest}
The authors declare that they have no conflict of interest.

\section*{Availability of data}
Not applicable.

\section*{Availability of code}
All codes implemented in this paper are available upon request from the corresponding author.

\begin{appendices}\label{app A}

\onecolumn
\section{Higher-order transverse discontinuity mapping}

For the phenomenological system comprising non-smooth fluid-structure interaction when subjected to impacts with a rigid barrier, the higher-order TDM \cite{chawla2022stability} maps perturbation vectors in the local neighbourhood of an impacting orbit from $\mathbf{x}_0$ to $\mathbf{x}_4$. Since, $\mathbf{x}_0 = \mathbf{x}_i + \mathbf{y}_-$ and $\mathbf{x}_4 = \mathbf{R}(\mathbf{x}_i) + \mathbf{y}_+$ in Fig. \ref{fig 1}, the TDM maps perturbations from $\mathbf{y}_-$ to $\mathbf{y}_+$ and is given by,
\begin{equation}\label{eq a1}
	\mathbf{y}_+ = \mathbf{A} + \mathbf{B} + \mathbf{C} + \mathbf{D} + \mathbf{E} + \mathbf{F} + \mathbf{G} + \mathbf{H},
\end{equation}
where,
\begin{align}
	\mathbf{A} &= \begin{pmatrix}
					y_1\\
					-r y_2\\
					y_3\\
					y_4
				  \end{pmatrix},\nonumber \\
	\mathbf{B} &= \delta_+ \begin{pmatrix}
							\dot{y}\\
							-r(\delta^2 y - \alpha \dot{y} + M x)\\
							\dot{x}\\
							- x - \epsilon(x^2 - 1)\dot{x} + A(\delta^2 y - \alpha \dot{y} + M x)
				  		   \end{pmatrix},\nonumber \\
	\mathbf{C} &= \delta_+ \begin{pmatrix}
							r \dot{y}\\
							-(- \delta^2 y + \alpha r \dot{y} + M x)\\
							-\dot{x}\\
							x + \epsilon(x^2 - 1)\dot{x} - A(- \delta^2 y + \alpha r \dot{y} + M x)
						   \end{pmatrix},\nonumber
\end{align}
\begin{align}
		\mathbf{D} &= \delta_+ \begin{pmatrix}
								y_2\\
								-r(- \delta^2 y_1 - \alpha y_2 + M y_3)\\
								y_4\\
								-A\delta^2 y_1 - A \alpha y_2 + (AM - 1 - 2\epsilon x \dot{x})y_3 - \epsilon(x^2 - 1) y_4
							   \end{pmatrix},\nonumber\\
		\mathbf{E} &= \frac{1}{2}\delta^2_+ \begin{pmatrix}
												-\delta^2 y - \alpha \dot{y} + M x\\
												-r(-\delta^2 \dot{y} - \alpha(-\delta^2 y - \alpha \dot{y} + M x) + M \dot{x})\\
												- x -\epsilon (x^2 - 1)\dot{x} + A(-\delta^2 y - \alpha \dot{y} + M x)\\
												-A\dot{y}\delta^2 - \alpha A(-\delta^2 y - \alpha \dot{y} + M x) + \dot{x}(A M - 1 - 2\epsilon x \dot{x}) \hdots \\
												- \epsilon(x^2 - 1)(-x - \epsilon(x^2 - 1)\dot{x} + A(-\delta^2 y - \alpha \dot{y} + M x))
					  						\end{pmatrix},\nonumber\\
		\mathbf{F} &= \delta_+ \begin{pmatrix}
								-r^2 y_2\\
								\delta^2 y_1 + \alpha r^2 y_2 - M y_3\\
								-y_4\\
								A\delta^2 y_1 + A \alpha r^2 y_2 - (AM - 1 - 2\epsilon x \dot{x}) + \epsilon(x^2 - 1)y_4
							   \end{pmatrix},\nonumber\\
		\mathbf{G} &= \delta_+^2 \begin{pmatrix}
									-r^2(-\delta^2 y - \alpha \dot{y} + M x)\\
									\delta^2 \dot{y} + \alpha r^2(-\delta^2 y - \alpha \dot{y} + M x) + M\dot{x}\\
									x + \epsilon(x^2 - 1)\dot{x} - A(-\delta^2 y - \alpha \dot{y} + M x)\\
									\delta^2 A \dot{y} + r^2 A \alpha(-\delta^2 y - \alpha \dot{y} + M x) - (AM - 1 - 2\epsilon x \dot{x})\dot{x} \hdots \\
									+ \epsilon(x^2 - 1)(-x - \epsilon(x^2 - 1)\dot{x} + A(-\delta^2 y - \alpha \dot{y} + M x)) 
								 \end{pmatrix},\nonumber\\
		\mathbf{H} &= \frac{1}{2}\delta_+^2 \begin{pmatrix}
												-r(-\delta^2 y + r \alpha \dot{y} + M x)\\
												r \delta^2 \dot{y} + r \alpha(-\delta^2 y + r \alpha \dot{y} + M x) + M\dot{x}\\
												-x - \epsilon(x^2 - 1)\dot{x} + A(-\delta^2 y + r \alpha \dot{y} + M x)\\
												r A \delta^2 \dot{y} + r A \alpha(-\delta^2 y + r \alpha \dot{y} + M x) + (AM - 1 - 2\epsilon x \dot{x})\dot{x}\hdots \\
												-\epsilon(x^2 - 1)(-x - \epsilon(x^2 - 1)\dot{x} + A(-\delta^2 y + r \alpha \dot{y} + M x))
								 			\end{pmatrix},\nonumber
\end{align}
$\alpha = (2 \xi \delta + \gamma/\mu)$ and $y_i$ are the components of the perturbed state $\mathbf{y}_-$.

\twocolumn
\end{appendices}

\bibliography{References}


\begin{thebibliography}{56}
\ifx \bisbn   \undefined \def \bisbn  #1{ISBN #1}\fi
\ifx \binits  \undefined \def \binits#1{#1}\fi
\ifx \bauthor  \undefined \def \bauthor#1{#1}\fi
\ifx \batitle  \undefined \def \batitle#1{#1}\fi
\ifx \bjtitle  \undefined \def \bjtitle#1{#1}\fi
\ifx \bvolume  \undefined \def \bvolume#1{\textbf{#1}}\fi
\ifx \byear  \undefined \def \byear#1{#1}\fi
\ifx \bissue  \undefined \def \bissue#1{#1}\fi
\ifx \bfpage  \undefined \def \bfpage#1{#1}\fi
\ifx \blpage  \undefined \def \blpage #1{#1}\fi
\ifx \burl  \undefined \def \burl#1{\textsf{#1}}\fi
\ifx \doiurl  \undefined \def \doiurl#1{\url{https://doi.org/#1}}\fi
\ifx \betal  \undefined \def \betal{\textit{et al.}}\fi
\ifx \binstitute  \undefined \def \binstitute#1{#1}\fi
\ifx \binstitutionaled  \undefined \def \binstitutionaled#1{#1}\fi
\ifx \bctitle  \undefined \def \bctitle#1{#1}\fi
\ifx \beditor  \undefined \def \beditor#1{#1}\fi
\ifx \bpublisher  \undefined \def \bpublisher#1{#1}\fi
\ifx \bbtitle  \undefined \def \bbtitle#1{#1}\fi
\ifx \bedition  \undefined \def \bedition#1{#1}\fi
\ifx \bseriesno  \undefined \def \bseriesno#1{#1}\fi
\ifx \blocation  \undefined \def \blocation#1{#1}\fi
\ifx \bsertitle  \undefined \def \bsertitle#1{#1}\fi
\ifx \bsnm \undefined \def \bsnm#1{#1}\fi
\ifx \bsuffix \undefined \def \bsuffix#1{#1}\fi
\ifx \bparticle \undefined \def \bparticle#1{#1}\fi
\ifx \barticle \undefined \def \barticle#1{#1}\fi
\bibcommenthead
\ifx \bconfdate \undefined \def \bconfdate #1{#1}\fi
\ifx \botherref \undefined \def \botherref #1{#1}\fi
\ifx \url \undefined \def \url#1{\textsf{#1}}\fi
\ifx \bchapter \undefined \def \bchapter#1{#1}\fi
\ifx \bbook \undefined \def \bbook#1{#1}\fi
\ifx \bcomment \undefined \def \bcomment#1{#1}\fi
\ifx \oauthor \undefined \def \oauthor#1{#1}\fi
\ifx \citeauthoryear \undefined \def \citeauthoryear#1{#1}\fi
\ifx \endbibitem  \undefined \def \endbibitem {}\fi
\ifx \bconflocation  \undefined \def \bconflocation#1{#1}\fi
\ifx \arxivurl  \undefined \def \arxivurl#1{\textsf{#1}}\fi
\csname PreBibitemsHook\endcsname

\bibitem{xue2015practical}
\begin{barticle}
\bauthor{\bsnm{Xue}, \binits{H.}},
\bauthor{\bsnm{Wang}, \binits{K.}},
\bauthor{\bsnm{Tang}, \binits{W.}}:
\batitle{A practical approach to predicting cross-flow and in-line viv response
  for deepwater risers}.
\bjtitle{Applied ocean research}
\bvolume{52},
\bfpage{92}--\blpage{101}
(\byear{2015})
\end{barticle}
\endbibitem

\bibitem{raiola2016wake}
\begin{barticle}
\bauthor{\bsnm{Raiola}, \binits{M.}},
\bauthor{\bsnm{Ianiro}, \binits{A.}},
\bauthor{\bsnm{Discetti}, \binits{S.}}:
\batitle{Wake of tandem cylinders near a wall}.
\bjtitle{Experimental Thermal and Fluid Science}
\bvolume{78},
\bfpage{354}--\blpage{369}
(\byear{2016})
\end{barticle}
\endbibitem

\bibitem{de2022vortex}
\begin{barticle}
\bauthor{\bsnm{De}, \binits{A.K.}},
\bauthor{\bsnm{Sarkar}, \binits{S.}}:
\batitle{Vortex induced vibration of a circular cylinder colliding with a rigid
  wall}.
\bjtitle{Physical Review Fluids}
\bvolume{7}(\bissue{6}),
\bfpage{064702}
(\byear{2022})
\end{barticle}
\endbibitem

\bibitem{hu2024model}
\begin{barticle}
\bauthor{\bsnm{Hu}, \binits{Y.}},
\bauthor{\bsnm{Qian}, \binits{Q.}},
\bauthor{\bsnm{Li}, \binits{L.}},
\bauthor{\bsnm{Hu}, \binits{X.}},
\bauthor{\bsnm{Mei}, \binits{X.}},
\bauthor{\bsnm{Li}, \binits{G.}},
\bauthor{\bsnm{Gu}, \binits{B.}}:
\batitle{Model and collision prevention control for 2-d vessel-riser array
  system based on improved wake interference model}.
\bjtitle{Ocean Engineering}
\bvolume{295},
\bfpage{116772}
(\byear{2024})
\end{barticle}
\endbibitem

\bibitem{Ellinas}
\begin{barticle}
\bauthor{\bsnm{Ellinas}, \binits{C.P.}},
\bauthor{\bsnm{Valsgard}, \binits{S.}}:
\batitle{{Collisions and Damage of Offshore Structures: A State-of-the-Art}}.
\bjtitle{Journal of Energy Resources Technology}
\bvolume{107}(\bissue{3}),
\bfpage{297}--\blpage{314}
(\byear{1985})
\end{barticle}
\endbibitem

\bibitem{tian2020static}
\begin{barticle}
\bauthor{\bsnm{Tian}, \binits{D.}},
\bauthor{\bsnm{Fan}, \binits{H.}},
\bauthor{\bsnm{Leira}, \binits{B.J.}},
\bauthor{\bsnm{S{\ae}vik}, \binits{S.}},
\bauthor{\bsnm{Fu}, \binits{P.}}:
\batitle{Static analysis of interaction between two adjacent top tensioned
  risers with consideration of wake effects}.
\bjtitle{Ocean Engineering}
\bvolume{195},
\bfpage{106662}
(\byear{2020})
\end{barticle}
\endbibitem

\bibitem{chandrasekaran2020design}
\begin{bbook}
\bauthor{\bsnm{Chandrasekaran}, \binits{S.}}:
\bbtitle{Design of Marine Risers with Functionally Graded Materials}.
\bpublisher{Woodhead Publishing}, \blocation{???}
(\byear{2020})
\end{bbook}
\endbibitem

\bibitem{huse2000impulse}
\begin{bchapter}
\bauthor{\bsnm{Huse}, \binits{E.}},
\bauthor{\bsnm{Kleiven}, \binits{G.}}:
\bctitle{Impulse and energy in deepsea riser collisions owing to wake
  interference}.
In: \bbtitle{Offshore Technology Conference},
p. \bfpage{11993}
(\byear{2000}).
\bcomment{OTC}
\end{bchapter}
\endbibitem

\bibitem{paidoussis1992cross}
\begin{barticle}
\bauthor{\bsnm{Pa{\"\i}doussis}, \binits{M.P.}},
\bauthor{\bsnm{Li}, \binits{G.X.}}:
\batitle{Cross-flow-induced chaotic vibrations of heat-exchanger tubes
  impacting on loose supports}.
\bjtitle{Journal of Sound and Vibration}
\bvolume{152}(\bissue{2}),
\bfpage{305}--\blpage{326}
(\byear{1992})
\end{barticle}
\endbibitem

\bibitem{virgin2009some}
\begin{bchapter}
\bauthor{\bsnm{Virgin}, \binits{L.}},
\bauthor{\bsnm{Plaut}, \binits{R.}}:
\bctitle{Some non-smooth dynamical systems in offshore mechanics}.
In: \bbtitle{Vibro-Impact Dynamics of Ocean Systems and Related Problems},
pp. \bfpage{259}--\blpage{268}.
\bpublisher{Springer}, \blocation{???}
(\byear{2009})
\end{bchapter}
\endbibitem

\bibitem{xue2023nonlinear}
\begin{bchapter}
\bauthor{\bsnm{Xue}, \binits{B.}},
\bauthor{\bsnm{Mao}, \binits{Y.}},
\bauthor{\bsnm{Zhang}, \binits{C.}},
\bauthor{\bsnm{Zhang}, \binits{H.}},
\bauthor{\bsnm{Qi}, \binits{Z.}}:
\bctitle{Nonlinear fluid-induced vibro-impact analysis on the fatigue failure
  pattern of a large-scale trashrack with a reduced-order model}.
In: \bbtitle{Structures},
vol. \bseriesno{49},
pp. \bfpage{467}--\blpage{478}
(\byear{2023}).
\bcomment{Elsevier}
\end{bchapter}
\endbibitem

\bibitem{ibrahim2014recent}
\begin{barticle}
\bauthor{\bsnm{Ibrahim}, \binits{R.A.}}:
\batitle{Recent advances in vibro-impact dynamics and collision of ocean
  vessels}.
\bjtitle{Journal of Sound and Vibration}
\bvolume{333}(\bissue{23}),
\bfpage{5900}--\blpage{5916}
(\byear{2014})
\end{barticle}
\endbibitem

\bibitem{williamson2004vortex}
\begin{barticle}
\bauthor{\bsnm{Williamson}, \binits{C.H.}},
\bauthor{\bsnm{Govardhan}, \binits{R.}}:
\batitle{Vortex-induced vibrations}.
\bjtitle{Annual Review of Fluid Mechanics}
\bvolume{36},
\bfpage{413}--\blpage{455}
(\byear{2004})
\end{barticle}
\endbibitem

\bibitem{williamson2008brief}
\begin{barticle}
\bauthor{\bsnm{Williamson}, \binits{C.}},
\bauthor{\bsnm{Govardhan}, \binits{R.}}:
\batitle{A brief review of recent results in vortex-induced vibrations}.
\bjtitle{Journal of Wind engineering and industrial Aerodynamics}
\bvolume{96}(\bissue{6-7}),
\bfpage{713}--\blpage{735}
(\byear{2008})
\end{barticle}
\endbibitem

\bibitem{bernardo2008piecewise}
\begin{botherref}
\oauthor{\bsnm{Bernardo}, \binits{M.}},
\oauthor{\bsnm{Budd}, \binits{C.}},
\oauthor{\bsnm{Champneys}, \binits{A.R.}},
\oauthor{\bsnm{Kowalczyk}, \binits{P.}}:
Piecewise-smooth dynamical systems: theory and applications.
Springer Science \& Business Media
\textbf{163}
(2008)
\end{botherref}
\endbibitem

\bibitem{nordmark1991non}
\begin{barticle}
\bauthor{\bsnm{Nordmark}, \binits{A.B.}}:
\batitle{Non-periodic motion caused by grazing incidence in an impact
  oscillator}.
\bjtitle{Journal of Sound and Vibration}
\bvolume{145}(\bissue{2}),
\bfpage{279}--\blpage{297}
(\byear{1991})
\end{barticle}
\endbibitem

\bibitem{chawla2022stability}
\begin{botherref}
\oauthor{\bsnm{Chawla}, \binits{R.}},
\oauthor{\bsnm{Rounak}, \binits{A.}},
\oauthor{\bsnm{Pakrashi}, \binits{V.}}:
Stability analysis of hybrid systems with higher order transverse discontinuity
  mapping.
arXiv preprint arXiv:2203.13222
(2022)
\end{botherref}
\endbibitem

\bibitem{muller1995calculation}
\begin{barticle}
\bauthor{\bsnm{M{\"u}ller}, \binits{P.C.}}:
\batitle{Calculation of lyapunov exponents for dynamic systems with
  discontinuities}.
\bjtitle{Chaos, Solitons \& Fractals}
\bvolume{5}(\bissue{9}),
\bfpage{1671}--\blpage{1681}
(\byear{1995})
\end{barticle}
\endbibitem

\bibitem{facchinetti2004coupling}
\begin{barticle}
\bauthor{\bsnm{Facchinetti}, \binits{M.L.}},
\bauthor{\bsnm{De~Langre}, \binits{E.}},
\bauthor{\bsnm{Biolley}, \binits{F.}}:
\batitle{Coupling of structure and wake oscillators in vortex-induced
  vibrations}.
\bjtitle{Journal of Fluids and structures}
\bvolume{19}(\bissue{2}),
\bfpage{123}--\blpage{140}
(\byear{2004})
\end{barticle}
\endbibitem

\bibitem{newman1997direct}
\begin{barticle}
\bauthor{\bsnm{Newman}, \binits{D.J.}},
\bauthor{\bsnm{Karniadakis}, \binits{G.E.}}:
\batitle{A direct numerical simulation study of flow past a freely vibrating
  cable}.
\bjtitle{Journal of Fluid Mechanics}
\bvolume{344},
\bfpage{95}--\blpage{136}
(\byear{1997})
\end{barticle}
\endbibitem

\bibitem{sarpkaya2004critical}
\begin{barticle}
\bauthor{\bsnm{Sarpkaya}, \binits{T.}}:
\batitle{A critical review of the intrinsic nature of vortex-induced
  vibrations}.
\bjtitle{Journal of fluids and structures}
\bvolume{19}(\bissue{4}),
\bfpage{389}--\blpage{447}
(\byear{2004})
\end{barticle}
\endbibitem

\bibitem{gabbai2005overview}
\begin{barticle}
\bauthor{\bsnm{Gabbai}, \binits{R.D.}},
\bauthor{\bsnm{Benaroya}, \binits{H.}}:
\batitle{An overview of modeling and experiments of vortex-induced vibration of
  circular cylinders}.
\bjtitle{Journal of sound and vibration}
\bvolume{282}(\bissue{3-5}),
\bfpage{575}--\blpage{616}
(\byear{2005})
\end{barticle}
\endbibitem

\bibitem{bishop1964lift}
\begin{barticle}
\bauthor{\bsnm{Bishop}, \binits{R.E.D.}},
\bauthor{\bsnm{Hassan}, \binits{A.}}:
\batitle{The lift and drag forces on a circular cylinder oscillating in a
  flowing fluid}.
\bjtitle{Proceedings of the Royal Society of London. Series A. Mathematical and
  Physical Sciences}
\bvolume{277}(\bissue{1368}),
\bfpage{51}--\blpage{75}
(\byear{1964})
\end{barticle}
\endbibitem

\bibitem{balasubramanian1996nonlinear}
\begin{barticle}
\bauthor{\bsnm{Balasubramanian}, \binits{S.}},
\bauthor{\bsnm{Skop}, \binits{R.}}:
\batitle{A nonlinear oscillator model for vortex shedding from cylinders and
  cones in uniform and shear flows}.
\bjtitle{Journal of Fluids and Structures}
\bvolume{10}(\bissue{3}),
\bfpage{197}--\blpage{214}
(\byear{1996})
\end{barticle}
\endbibitem

\bibitem{skop1997new}
\begin{barticle}
\bauthor{\bsnm{Skop}, \binits{R.A.}},
\bauthor{\bsnm{Balasubramanian}, \binits{S.}}:
\batitle{A new twist on an old model for vortex-excited vibrations}.
\bjtitle{Journal of Fluids and Structures}
\bvolume{11}(\bissue{4}),
\bfpage{395}--\blpage{412}
(\byear{1997})
\end{barticle}
\endbibitem

\bibitem{hartlen1970lift}
\begin{barticle}
\bauthor{\bsnm{Hartlen}, \binits{R.T.}},
\bauthor{\bsnm{Currie}, \binits{I.G.}}:
\batitle{Lift-oscillator model of vortex-induced vibration}.
\bjtitle{Journal of the Engineering Mechanics Division}
\bvolume{96}(\bissue{5}),
\bfpage{577}--\blpage{591}
(\byear{1970})
\end{barticle}
\endbibitem

\bibitem{krenk1999energy}
\begin{barticle}
\bauthor{\bsnm{Krenk}, \binits{S.}},
\bauthor{\bsnm{Nielsen}, \binits{S.R.}}:
\batitle{Energy balanced double oscillator model for vortex-induced
  vibrations}.
\bjtitle{Journal of Engineering Mechanics}
\bvolume{125}(\bissue{3}),
\bfpage{263}--\blpage{271}
(\byear{1999})
\end{barticle}
\endbibitem

\bibitem{ogink2010wake}
\begin{barticle}
\bauthor{\bsnm{Ogink}, \binits{R.}},
\bauthor{\bsnm{Metrikine}, \binits{A.}}:
\batitle{A wake oscillator with frequency dependent coupling for the modeling
  of vortex-induced vibration}.
\bjtitle{Journal of sound and vibration}
\bvolume{329}(\bissue{26}),
\bfpage{5452}--\blpage{5473}
(\byear{2010})
\end{barticle}
\endbibitem

\bibitem{qu2020single}
\begin{barticle}
\bauthor{\bsnm{Qu}, \binits{Y.}},
\bauthor{\bsnm{Metrikine}, \binits{A.V.}}:
\batitle{A single van der pol wake oscillator model for coupled cross-flow and
  in-line vortex-induced vibrations}.
\bjtitle{Ocean Engineering}
\bvolume{196},
\bfpage{106732}
(\byear{2020})
\end{barticle}
\endbibitem

\bibitem{mureithi2000bifurcation}
\begin{bchapter}
\bauthor{\bsnm{Mureithi}, \binits{N.}},
\bauthor{\bsnm{Kanki}, \binits{H.}},
\bauthor{\bsnm{Nakamura}, \binits{T.}}:
\bctitle{Bifurcation and perturbation analysis of some vortex shedding models}.
In: \bbtitle{Proceedings of the Seventh International Conference on
  Flow-Induced Vibrations, Luzern, Switzerland. Balkema, Rotterdam},
pp. \bfpage{61}--\blpage{68}
(\byear{2000})
\end{bchapter}
\endbibitem

\bibitem{plaschko2000global}
\begin{barticle}
\bauthor{\bsnm{Plaschko}, \binits{P.}}:
\batitle{Global chaos in flow-induced oscillations of cylinders}.
\bjtitle{Journal of Fluids and Structures}
\bvolume{14}(\bissue{6}),
\bfpage{883}--\blpage{893}
(\byear{2000})
\end{barticle}
\endbibitem

\bibitem{govardhan2000modes}
\begin{barticle}
\bauthor{\bsnm{Govardhan}, \binits{R.}},
\bauthor{\bsnm{Williamson}, \binits{C.}}:
\batitle{Modes of vortex formation and frequency response of a freely vibrating
  cylinder}.
\bjtitle{Journal of Fluid Mechanics}
\bvolume{420},
\bfpage{85}--\blpage{130}
(\byear{2000})
\end{barticle}
\endbibitem

\bibitem{govardhan2004critical}
\begin{barticle}
\bauthor{\bsnm{Govardhan}, \binits{R.}},
\bauthor{\bsnm{Williamson}, \binits{C.}}:
\batitle{Critical mass in vortex-induced vibration of a cylinder}.
\bjtitle{European Journal of Mechanics-B/Fluids}
\bvolume{23}(\bissue{1}),
\bfpage{17}--\blpage{27}
(\byear{2004})
\end{barticle}
\endbibitem

\bibitem{khalak1999motions}
\begin{barticle}
\bauthor{\bsnm{Khalak}, \binits{A.}},
\bauthor{\bsnm{Williamson}, \binits{C.H.}}:
\batitle{Motions, forces and mode transitions in vortex-induced vibrations at
  low mass-damping}.
\bjtitle{Journal of fluids and Structures}
\bvolume{13}(\bissue{7-8}),
\bfpage{813}--\blpage{851}
(\byear{1999})
\end{barticle}
\endbibitem

\bibitem{banerjee2009invisible}
\begin{barticle}
\bauthor{\bsnm{Banerjee}, \binits{S.}},
\bauthor{\bsnm{Ing}, \binits{J.}},
\bauthor{\bsnm{Pavlovskaia}, \binits{E.}},
\bauthor{\bsnm{Wiercigroch}, \binits{M.}},
\bauthor{\bsnm{Reddy}, \binits{R.K.}}:
\batitle{Invisible grazings and dangerous bifurcations in impacting systems:
  the problem of narrow-band chaos}.
\bjtitle{Physical Review E}
\bvolume{79}(\bissue{3}),
\bfpage{037201}
(\byear{2009})
\end{barticle}
\endbibitem

\bibitem{di2002bifurcations}
\begin{barticle}
\bauthor{\bsnm{Di~Bernardo}, \binits{M.}},
\bauthor{\bsnm{Kowalczyk}, \binits{P.}},
\bauthor{\bsnm{Nordmark}, \binits{A.}}:
\batitle{Bifurcations of dynamical systems with sliding: derivation of
  normal-form mappings}.
\bjtitle{Physica D: Nonlinear Phenomena}
\bvolume{170}(\bissue{3-4}),
\bfpage{175}--\blpage{205}
(\byear{2002})
\end{barticle}
\endbibitem

\bibitem{budd1994chattering}
\begin{barticle}
\bauthor{\bsnm{Budd}, \binits{C.}},
\bauthor{\bsnm{Dux}, \binits{F.}}:
\batitle{Chattering and related behaviour in impact oscillators}.
\bjtitle{Philosophical Transactions of the Royal Society of London. Series A:
  Physical and Engineering Sciences}
\bvolume{347}(\bissue{1683}),
\bfpage{365}--\blpage{389}
(\byear{1994})
\end{barticle}
\endbibitem

\bibitem{nusse1994border}
\begin{barticle}
\bauthor{\bsnm{Nusse}, \binits{H.E.}},
\bauthor{\bsnm{Ott}, \binits{E.}},
\bauthor{\bsnm{Yorke}, \binits{J.A.}}:
\batitle{Border-collision bifurcations: An explanation for observed bifurcation
  phenomena}.
\bjtitle{Physical Review E}
\bvolume{49}(\bissue{2}),
\bfpage{1073}
(\byear{1994})
\end{barticle}
\endbibitem

\bibitem{chin1994grazing}
\begin{barticle}
\bauthor{\bsnm{Chin}, \binits{W.}},
\bauthor{\bsnm{Ott}, \binits{E.}},
\bauthor{\bsnm{Nusse}, \binits{H.E.}},
\bauthor{\bsnm{Grebogi}, \binits{C.}}:
\batitle{Grazing bifurcations in impact oscillators}.
\bjtitle{Physical Review E}
\bvolume{50}(\bissue{6}),
\bfpage{4427}
(\byear{1994})
\end{barticle}
\endbibitem

\bibitem{jiang2017grazing}
\begin{barticle}
\bauthor{\bsnm{Jiang}, \binits{H.}},
\bauthor{\bsnm{Chong}, \binits{A.S.}},
\bauthor{\bsnm{Ueda}, \binits{Y.}},
\bauthor{\bsnm{Wiercigroch}, \binits{M.}}:
\batitle{Grazing-induced bifurcations in impact oscillators with elastic and
  rigid constraints}.
\bjtitle{International Journal of Mechanical Sciences}
\bvolume{127},
\bfpage{204}--\blpage{214}
(\byear{2017})
\end{barticle}
\endbibitem

\bibitem{oestreich1996bifurcation}
\begin{barticle}
\bauthor{\bsnm{Oestreich}, \binits{M.}},
\bauthor{\bsnm{Hinrichs}, \binits{N.}},
\bauthor{\bsnm{Popp}, \binits{K.}}:
\batitle{Bifurcation and stability analysis for a non-smooth friction
  oscillator}.
\bjtitle{Archive of Applied Mechanics}
\bvolume{66}(\bissue{5}),
\bfpage{301}--\blpage{314}
(\byear{1996})
\end{barticle}
\endbibitem

\bibitem{awrejcewicz2003bifurcation}
\begin{botherref}
\oauthor{\bsnm{Awrejcewicz}, \binits{J.}},
\oauthor{\bsnm{Lamarque}, \binits{C.-H.}}:
Bifurcation and chaos in nonsmooth mechanical systems.
World Scientific
\textbf{45}
(2003)
\end{botherref}
\endbibitem

\bibitem{nayfeh2008applied}
\begin{bbook}
\bauthor{\bsnm{Nayfeh}, \binits{A.H.}},
\bauthor{\bsnm{Balachandran}, \binits{B.}}:
\bbtitle{Applied Nonlinear Dynamics: Analytical, Computational, and
  Experimental Methods}.
\bpublisher{John Wiley \& Sons}, \blocation{???}
(\byear{2008})
\end{bbook}
\endbibitem

\bibitem{stefanski2000estimation}
\begin{barticle}
\bauthor{\bsnm{Stefanski}, \binits{A.}}:
\batitle{Estimation of the largest lyapunov exponent in systems with impacts}.
\bjtitle{Chaos, Solitons \& Fractals}
\bvolume{11}(\bissue{15}),
\bfpage{2443}--\blpage{2451}
(\byear{2000})
\end{barticle}
\endbibitem

\bibitem{balcerzak2020determining}
\begin{barticle}
\bauthor{\bsnm{Balcerzak}, \binits{M.}},
\bauthor{\bsnm{Dabrowski}, \binits{A.}},
\bauthor{\bsnm{Blazejczyk--Okolewska}, \binits{B.}},
\bauthor{\bsnm{Stefanski}, \binits{A.}}:
\batitle{Determining lyapunov exponents of non-smooth systems: Perturbation
  vectors approach}.
\bjtitle{Mechanical Systems and Signal Processing}
\bvolume{141},
\bfpage{106734}
(\byear{2020})
\end{barticle}
\endbibitem

\bibitem{leine2012non}
\begin{barticle}
\bauthor{\bsnm{Leine}, \binits{R.}}:
\batitle{Non-smooth stability analysis of the parametrically excited impact
  oscillator}.
\bjtitle{International Journal of Non-Linear Mechanics}
\bvolume{47}(\bissue{9}),
\bfpage{1020}--\blpage{1032}
(\byear{2012})
\end{barticle}
\endbibitem

\bibitem{coleman1997motions}
\begin{barticle}
\bauthor{\bsnm{Coleman}, \binits{M.J.}},
\bauthor{\bsnm{Chatterjee}, \binits{A.}},
\bauthor{\bsnm{Ruina}, \binits{A.}}:
\batitle{Motions of a rimless spoked wheel: a simple three-dimensional system
  with impacts}.
\bjtitle{Dynamics and stability of systems}
\bvolume{12}(\bissue{3}),
\bfpage{139}--\blpage{159}
(\byear{1997})
\end{barticle}
\endbibitem

\bibitem{yin2018higher}
\begin{barticle}
\bauthor{\bsnm{Yin}, \binits{S.}},
\bauthor{\bsnm{Wen}, \binits{G.}},
\bauthor{\bsnm{Xu}, \binits{H.}},
\bauthor{\bsnm{Wu}, \binits{X.}}:
\batitle{Higher order zero time discontinuity mapping for analysis of
  degenerate grazing bifurcations of impacting oscillators}.
\bjtitle{Journal of Sound and Vibration}
\bvolume{437},
\bfpage{209}--\blpage{222}
(\byear{2018})
\end{barticle}
\endbibitem

\bibitem{blevins1977flow}
\begin{botherref}
\oauthor{\bsnm{Blevins}, \binits{R.D.}}:
Flow-induced vibration.
New York
(1977)
\end{botherref}
\endbibitem

\bibitem{king1977vortex}
\begin{botherref}
\oauthor{\bsnm{King}, \binits{R.}}:
Vortex excited oscillations of yawed circular cylinders
(1977)
\end{botherref}
\endbibitem

\bibitem{griffin1980vortex}
\begin{botherref}
\oauthor{\bsnm{Griffin}, \binits{O.}}:
Vortex-excited cross-flow vibrations of a single cylindrical tube
(1980)
\end{botherref}
\endbibitem

\bibitem{pantazopoulos1994vortex}
\begin{botherref}
\oauthor{\bsnm{Pantazopoulos}, \binits{M.S.}}:
Vortex-induced vibration parameters: critical review
(1994)
\end{botherref}
\endbibitem

\bibitem{vickery1964flow}
\begin{bchapter}
\bauthor{\bsnm{Vickery}, \binits{B.}},
\bauthor{\bsnm{Watkins}, \binits{R.}}:
\bctitle{Flow-induced vibrations of cylindrical structures}.
In: \bbtitle{Hydraulics and Fluid Mechanics},
pp. \bfpage{213}--\blpage{241}.
\bpublisher{Elsevier}, \blocation{???}
(\byear{1964})
\end{bchapter}
\endbibitem

\bibitem{bearman1984vortex}
\begin{barticle}
\bauthor{\bsnm{Bearman}, \binits{P.W.}}:
\batitle{Vortex shedding from oscillating bluff bodies}.
\bjtitle{Annual review of fluid mechanics}
\bvolume{16}(\bissue{1}),
\bfpage{195}--\blpage{222}
(\byear{1984})
\end{barticle}
\endbibitem

\bibitem{carberry2001forces}
\begin{barticle}
\bauthor{\bsnm{Carberry}, \binits{J.}},
\bauthor{\bsnm{Sheridan}, \binits{J.}},
\bauthor{\bsnm{Rockwell}, \binits{D.}}:
\batitle{Forces and wake modes of an oscillating cylinder}.
\bjtitle{Journal of Fluids and Structures}
\bvolume{15}(\bissue{3-4}),
\bfpage{523}--\blpage{532}
(\byear{2001})
\end{barticle}
\endbibitem

\bibitem{grassberger1983measuring}
\begin{barticle}
\bauthor{\bsnm{Grassberger}, \binits{P.}},
\bauthor{\bsnm{Procaccia}, \binits{I.}}:
\batitle{Measuring the strangeness of strange attractors}.
\bjtitle{Physica D: nonlinear phenomena}
\bvolume{9}(\bissue{1-2}),
\bfpage{189}--\blpage{208}
(\byear{1983})
\end{barticle}
\endbibitem

\end{thebibliography}

\end{document}